\documentclass[a4paper,12pt]{article}
\def\mydate{March 30, 2021}

\usepackage{graphicx}

\usepackage{amsmath}        
\usepackage{amssymb}        
\usepackage{cite}
\usepackage{cancel}
\usepackage{color}
\usepackage{colortbl}
\usepackage{multicol}
\usepackage{multirow}
\usepackage{lscape}
\usepackage[all,knot]{xy}
\usepackage{longtable}
\usepackage{braket}
\usepackage{bm}
\usepackage{fancybox}

\makeatletter
 
 \@addtoreset{equation}{section}
\makeatother

\newcommand{\SM}{\text{\rm SM}}

\def\ignore#1{{}}

\def\go{\rightarrow}

\def\SM{{\rm SM}}
\def\KK{{\rm KK}}

\def\la{\langle}
\def\ra{\rangle}

\def\mybig{\displaystyle \strut }

\def\myfrac#1#2{\frac{\mybig #1}{\mybig #2}}

\def\mymat#1#2{\begin{matrix}#1 \cr \noalign{\kern -2pt} #2\end{matrix}}

\def\mynoalign{\noalign{\kern 4pt}}
\def\mysnoalign{\noalign{\kern 3pt}}

\begin{document}
\allowdisplaybreaks

\thispagestyle{empty}


\leftline{\mydate  \hfill OU-HET-1094}
\rightline{KYUSHU-HET-222}

\vskip3.0cm

\baselineskip=30pt plus 1pt minus 1pt

\begin{center}

{\LARGE \bf Linear Collider Signals of $Z'$ Bosons} 

{\LARGE \bf in GUT Inspired Gauge-Higgs Unification}
\footnote[1]{Talk presented at the International Workshop on Future Linear Colliders (LCWS2021), 15-18 March 2021. C21-03-15.1.}

\end{center}


\baselineskip=22pt plus 1pt minus 1pt

\vskip 2.0cm

\begin{center}
{\bf Shuichiro Funatsu$^1$, Hisaki Hatanaka$^2$, Yutaka Hosotani$^3$,}

{\bf Yuta Orikasa$^4$ and Naoki Yamatsu$^5$\footnote[2]{Speaker}}

\baselineskip=17pt plus 1pt minus 1pt

\vskip 10pt
{\small \it $^1$Institute of Particle Physics and Key Laboratory of Quark and Lepton 
Physics (MOE), Central China Normal University, Wuhan, Hubei 430079, China} \\
{\small \it $^2$Osaka, Osaka 536-0014, Japan} \\
{\small \it $^3$Department of Physics, Osaka University, 
Toyonaka, Osaka 560-0043, Japan} \\
{\small \it $^4$Institute of Experimental and Applied Physics, Czech Technical University in Prague,} \\
{\small \it Husova 240/5, 110 00 Prague 1, Czech Republic} \\
{\small \it $^5$Department of Physics, Kyushu University, Fukuoka 819-0395, Japan} \\

\end{center}


\vskip 2.0cm

\begin{abstract}
 In gauge-Higgs unification (GHU), the 4D Higgs boson appears as a part
 of the fifth dimensional component of 5D gauge field. Recently, an
 $SO(11)$ GUT inspired $SO(5)\times U(1)\times SU(3)$ GHU model has
 been proposed. In the GHU, Kaluza-Klein (KK) excited states of neutral
 vector bosons, photon, $Z$ boson and $Z_R$ boson, appear as neutral
 massive vector bosons $Z'$s. The $Z'$ bosons in the GHU couple to
 quarks and  leptons with large parity violation, which leads to
 distinctive  polarization dependence in, e.g., cross sections and
 forward-backward  asymmetries in $e^-e^+\to\mu^-\mu^+,q\bar{q}$
 processes.  
 In the talk,
 we discuss fermion pair production in $e^-e^+$ linear
 collider experiments with polarized $e^-$ and $e^+$ beams in the GUT
 inspired GHU. Deviations from the SM are shown in the early stage of
 planned international linear collider (ILC) with 250 GeV
 experiments. The deviations can be tested for the  KK mass  scale up to
 about 15 TeV. 
 This talk is mainly based on Phys.Rev.D102(2020)015029
 (Ref.~\cite{Funatsu:2020haj}). 
\end{abstract}

\newpage

\baselineskip=20pt plus 1pt minus 1pt
\parskip=0pt

\section{Introduction}

The standard model (SM), $SU(3)_C\times SU(2)_L\times U(1)_Y$ gauge
theory, has been firmly established at low energies.
Since the SM is not the final theory of the nature from theoretical and 
experimental sides, there are many attempts for constructing models
beyond the SM by using many ideas such as grand unification,
supersymmetry, extra-dimension, etc. 

One attempt for constructing such a model is to consider gauge-Higgs
unification (GHU)
\cite{Hosotani:1983xw,Hosotani:1988bm,Davies:1987ei,Davies:1988wt,Hatanaka:1998yp,Hatanaka:1999sx}
in the framework of five or higher dimensional spacetime.
In GHU, the Higgs field appears as a fluctuation mode of the
Aharonov-Bohm  (AB) phase $\theta_H$ in the extra dimension.
The Higgs boson mass against quantum corrections is stabilized by the
gauge symmetry.
The $SU(3)_C \times SO(5) \times U(1)_X$ gauge theory in the
Randall-Sundrum (RS) warped space has been proposed in  
Refs.~\cite{Agashe:2004rs,Medina:2007hz,Hosotani:2008tx,Funatsu:2013ni,Funatsu:2014fda,Funatsu:2016uvi,Funatsu:2019xwr,Funatsu:2019fry,Funatsu:2020znj}. 
It gives nearly the same phenomenology at low energies as the SM
\cite{Funatsu:2013ni,Funatsu:2014fda,Funatsu:2015xba,Funatsu:2016uvi}. 
For example, deviations of the gauge couplings of quarks and leptons
from the SM values are less than 0.1\% for $\theta_H \simeq 0.1$. 
Higgs couplings of quarks, leptons, $W$ and $Z$ bosons are approximately 
the SM values times $\cos \theta_H$; the deviation is about 1{\%}.

The $SU(3)_C \times SO(5) \times U(1)_X$ GHU models predict $Z'$ bosons,
which are the  Kaluza-Klein (KK) modes of $\gamma$, $Z$, and $Z_R$.
In a GHU model, which is referred to as the {\it A-model} below,
quark-lepton multiplets are introduced in the vector representation of
$SO(5)$. Some results in the A-model
\cite{Funatsu:2014fda,Funatsu:2016uvi,Funatsu:2017nfm,Yoon:2018xud,Yoon:2018vsc,Funatsu:2019ujy,Hosotani:2019cnn,Funatsu:2020xeu}
are summarized below.
\begin{itemize}
\item The masses of $Z'$ bosons are in the 6\,TeV--9\,TeV range for
      $\theta_H\simeq 0.11$--0.07, which corresponds to the KK mass  
      scale $m_{\rm KK}\simeq$ 8\,TeV--11\,TeV.
      The current non-observation of $Z'$ signals at 14TeV Large Hadron
      Collider (LHC) puts the limit $\theta_H \lesssim 0.11$
      \cite{Funatsu:2014fda,Funatsu:2016uvi}.
\item Large parity violation appears in the couplings of quarks and
      leptons to KK gauge bosons, particularly to the $Z'$ bosons.
      Right-handed quarks and charged leptons have rather large
      couplings to $Z'$ bosons
      \cite{Funatsu:2017nfm,Yoon:2018xud,Funatsu:2019ujy}.
\item Effects of KK excited neutral vector bosons, $Z'$
      bosons, on $e^-e^+\to q\bar{q},\ell^-\ell^+$ cross sections are
      studied
      \cite{Funatsu:2017nfm,Yoon:2018xud,Funatsu:2019ujy}.
      In the process $e^-e^+ \go \mu^-\mu^+$, the deviation from the SM
      have large polarization dependence.
      The masses up to about 10TeV can be explored at 250 GeV
      International Linear Collider (ILC) with
      250 fb${}^{-1}$ data.
      (For other studies, see Refs.~\cite{Bilokin:2017lco,Richard:2018zhl,Irles:2019xny,Irles:2020gjh,Fujii:2017vwa,Aihara:2019gcq,Bambade:2019fyw}.)
\end{itemize}

Recently, we proposed another $SU(3)_C\times SO(5)\times U(1)_X$ GHU
model in Ref.~\cite{Funatsu:2019xwr}, which is referred to as the {\it
B-model} below.
Quark-lepton multiplets are introduced in the spinor, vector, and
singlet representations of $SO(5)$. The B-model can be embedded in the
$SO(11)$ gauge-Higgs grand unification
\cite{Hosotani:2015hoa,Yamatsu:2015rge,Furui:2016owe,Hosotani:2017krs,Hosotani:2017edv,Hosotani:2017ghg,Englert:2019xhz,Englert:2020eep},
where the SM gauge group and quark-lepton content are incorporated into
grand unified theory (GUT)
\cite{Georgi:1974sy,Inoue:1977qd,Fritzsch:1974nn,Gursey:1975ki,Slansky:1981yr,Yamatsu:2015gut}
in higher dimensional framework
\cite{Kawamura:1999nj,Kawamura:2000ir,Kawamura:2000ev,Burdman:2002se,Lim:2007jv,Kojima:2011ad,Kojima:2016fvv,Yamatsu:2017sgu,Yamatsu:2017ssg,Yamatsu:2018fsg,Yamatsu:2020usp,Maru:2019lit}.

In this talk, we mainly focus on observables in the
$e^-e^+\to \mu^-\mu^+$ process, i.e., cross section,
forward-backward asymmetry \cite{Schrempp:1987zy,Kennedy:1988rt},
left-right asymmetry
\cite{Schrempp:1987zy,Kennedy:1988rt,Abe:1994wx,Abe:1996nj}, 
and left-right forward-backward asymmetry
\cite{Blondel:1987gp,Kennedy:1988rt,Abe:1994bj,Abe:1994bm,Abe:1995yh}.
In Ref.~\cite{Funatsu:2020haj}, we also studied
$e^-e^+\to f\bar{f}(=c\bar{c},b\bar{b},t\bar{t})$ processes.
(For $Z'$ bosons in other models, see e.g.,
Refs.~\cite{Langacker:2008yv,Deguchi:2019tvp,Fujii:2019zll}.)

The proceeding is organized as follows.  
In Sec.~\ref{Sec:GHU_model}, the model is mention a bit.
In Sec.~\ref{Sec:Observables}, we quickly check observables such as
cross section, forward-backward, left-right, and left-right
forward-backward asymmetries.
In Sec.~\ref{Sec:Setup}, we describe how to detemine the parameter
sets in the model.
In Sec.~\ref{Sec:Reults}, we evaluate cross sections and other
observables in  $e^- e^+ \to \mu^-\mu^+$ process.
Section~\ref{Sec:Summary} is devoted to summary.

\section{GUT inspired GHU (B-model)}
\label{Sec:GHU_model}

The GUT inspired $SU(3)_C \times SO(5)\times U(1)_X$ GHU model
\cite{Funatsu:2019xwr} is defined in the RS warped space with metric 
\begin{align}
 ds^2= g_{MN} dx^M dx^N =e^{-2\sigma(y)} \eta_{\mu\nu}dx^\mu dx^\nu+dy^2,
\end{align} 
where $M,N=0,1,2,3,5$, $\mu,\nu=0,1,2,3$, $y=x^5$,
$\eta_{\mu\nu}=\mbox{diag}(-1,+1,+1,+1)$,
$\sigma(y)=\sigma(y+ 2L)=\sigma(-y)$,
and $\sigma(y)=ky$ for $0 \le y \le L$.
In terms of the conformal coordinate $z=e^{ky}$
($1\leq z\leq z_L=e^{kL}$) in the region $0 \leq y \leq L$ 
\begin{align}
ds^2= \frac{1}{z^2}
\bigg(\eta_{\mu\nu}dx^{\mu} dx^{\nu} + \frac{dz^2}{k^2}\bigg).
\end{align} 
The bulk region $0<y<L$ ($1<z<z_L$) is anti-de Sitter (AdS) spacetime 
with a cosmological constant $\Lambda=-6k^2$, which is sandwiched by the
UV brane at $y=0$ ($z=1$) and the IR brane at $y=L$ ($z=z_L$).  
The KK mass scale is $m_{\rm KK}=\pi k/(z_L-1) \simeq \pi kz_L^{-1}$
for $z_L\gg 1$.

Gauge fields of $SU(3)_C$, $SO(5)$, and $U(1)_X$ are denoted as
$A_M^{SU(3)_C}$, $A_M^{SO(5)}$, and $A_M^{U(1)_X}$, respectively.
The orbifold boundary conditions (BCs) are given by
\begin{align}
&\begin{pmatrix} A_\mu \cr  A_{y} \end{pmatrix} (x,y_j-y) =
P_{j} \begin{pmatrix} A_\mu \cr - A_{y} \end{pmatrix} (x,y_j+y)P_{j}^{-1},
\ \ (y_0,y_1=0,L)
 \label{Eq:BC-gauge}
\end{align}
for each gauge field.
$P_0=P_1= P_{\bf 3}^{SU(3)}=I_3$ for  $A_M^{SU(3)_C}$  and 
$P_0=P_1= 1$ for $A_M^{U(1)_X}$.  
$P_0=P_1 = P_{\bf 5}^{SO(5)}=\mbox{diag}\left(I_{4},-I_{1}\right)$
for $A_M^{SO(5)}$ in the vector representation.
The orbifold BCs $P_{\bf 4}^{SO(5)}$ and $P_{\bf 5}^{SO(5)}$ break
$SO(5)$ to $SO(4) \simeq SU(2)_L \times SU(2)_R$.
$W$, $Z$ bosons and $\gamma$ are zero modes in the $SO(4)$ part
of $A_\mu^{SO(5)}$, whereas the 4D Higgs boson is a zero mode in the 
$SO(5)/SO(4)$ part of $A_y^{SO(5)}$.
In the GHU model, extra neutral gauge bosons $Z'$ correspond to
KK photons $\gamma^{(n)}$,  KK $Z$ bosons $Z^{(n)}$,
and   KK $Z_R$ bosons $Z_R^{(n)}$ ($n \ge 1$),
where the $\gamma$, and $Z$, $Z_R$ bosons are the mass constants of
the electro-magnetic $U(1)_{\rm EM}$ neutral gauge bosons of
$SU(2)_L$, $SU(2)_R$, and $U(1)_X$.

Matter fields in the B-model are introduced both in the 5D bulk and on
the UV brane. 
They are listed in Table~\ref{Tab:matter}, which also contains the matter
content in the A-model for a reference.
The SM quark and lepton multiples are identified with 
the zero modes of the quark and lepton multiplets
$\Psi_{({\bf 3,4})}^{\alpha}$ $(\alpha=1,2,3)$,
$\Psi_{({\bf 3,1})}^{\pm \alpha}$, and
$\Psi_{({\bf 1,4})}^{\alpha}$ with appropriate BCs.
(For more information, see
Refs.~\cite{Funatsu:2019xwr,Funatsu:2019fry,Funatsu:2020znj}.)

\begin{table}[tbh]
\begin{center}
\begin{tabular}{|c|c|c|}
\hline
 \rowcolor[gray]{0.9}
&{B-model} &{A-model}\\
\hline
Quark
 &$({\bf 3}, {\bf 4})_{\frac{1}{6}}\ \ 
   ({\bf 3}, {\bf 1})_{-\frac{1}{3}}^+ \ \ 
   ({\bf 3}, {\bf 1})_{-\frac{1}{3}}^-$
 &$({\bf 3}, {\bf 5})_{\frac{2}{3}}\ \
   ({\bf 3}, {\bf 5})_{-\frac{1}{3}}$ \\
Lepton
 &$\strut ({\bf 1}, {\bf 4})_{-\frac{1}{2}}$ 
 &$({\bf 1}, {\bf 5})_{0} ~ ({\bf 1}, {\bf 5})_{-1}$  \\
\hline
Dark fermion
 & $({\bf 3}, {\bf 4})_{\frac{1}{6}}\ \
    ({\bf 1}, {\bf 5})_{0}^+ \ \
    ({\bf 1}, {\bf 5})_{0}^-$ 
 &$({\bf 1}, {\bf 4})_{\frac{1}{2}}$ \\
\hline
Brane fermion
 &$({\bf 1}, {\bf 1})_{0} $ 
 &$\begin{matrix} ({\bf 3}, [{\bf 2,1}])_{\frac{7}{6}, \frac{1}{6}, -\frac{5}{6}} \cr
({\bf 1}, [{\bf 2,1}])_{\frac{1}{2}, -\frac{1}{2}, -\frac{3}{2}} \end{matrix}$\\
\hline
Brane scalar &$({\bf 1}, {\bf 4})_{\frac{1}{2}} $ 
&$({\bf 1}, [{\bf 1,2}])_{\frac{1}{2}}$ \\
\hline
\end{tabular}
\caption{{\small
The $SU(3)_C\times SO(5) \times U(1)_X$ content of matter fields  is shown
in the  GUT inspired model (B-model) and the previous model (A-model).}
}
\label{Tab:matter}
\end{center}
\end{table}

The brane scalar field $\Phi_{({\bf 1}, {\bf 4})}(x)$
in $({\bf 1,4})_{\frac{1}{2}}$ of $SU(3)_C\times SO(5)\times U(1)_X$ in
Table~\ref{Tab:matter} is responsible for breaking $SO(5)\times U(1)_X$
to $SU(2)_L\times U(1)_Y$.
The nonvanishing vacuum expectation value (VEV) reduces the symmetry
$SU(3)_C\times SO(4) \times U(1)_X$ to 
the SM gauge group  $G_{\rm SM}\equiv SU(3)_C\times SU(2)_L\times U(1)_Y$.

The $U(1)_Y$ gauge boson is a mixed state of $U(1)_R(\subset SU(2)_R)$
and $U(1)_X$ gauge bosons. The $U(1)_Y$ gauge field $B_M^Y$ is
given in terms of the $SU(2)_R$ gauge fields $A_M^{a_R}$
$(a_R=1_R,2_R,3_R)$ and the $U(1)_X$ gauge field $B_M$  by 
$B_M^Y = s_\phi A_M^{3_R} + c_\phi  B_M$.
The mixing angle $\phi$ between $U(1)_R$ and $U(1)_X$ is given by 
$c_\phi = \cos \phi \equiv {g_A}/{\sqrt{g_A^2+g_B^2}}$ and
$s_\phi = \sin \phi \equiv {g_B}/{\sqrt{g_A^2+g_B^2}}$ where 
$g_A$ and $g_B$ are gauge couplings in $SO(5)$ and $U(1)_X$, respectively. 
The 4D $SU(2)_L$ gauge coupling is given by $g_w = g_A/\sqrt{L}$.  
The 5D gauge coupling $g_Y^{\rm 5D}$ of $U(1)_{Y}$
is given by $g_Y^{\rm 5D} ={g_Ag_B}/{\sqrt{g_A^2+g_B^2}}$.
The 4D bare Weinberg angle at the tree level $\theta_W^0$ is given by
\begin{align}
\sin \theta_W^0 = \frac{s_\phi}{\sqrt{\smash[b]{1 + s_\phi^2}}}.
\label{Eq:gY-sW}
\end{align}

The 4D Higgs boson $\phi_H(x)$ is the zero mode in the
$A_z = (kz)^{-1} A_y$ component:
\begin{align}
A_z^{(j5)} (x, z) &= \frac{1}{\sqrt{k}} \, \phi_j (x) u_H (z) + \cdots,\
u_H (z) = \sqrt{ \frac{2}{z_L^2 -1} } \, z ~,\ 
\phi_H(x) = \frac{1}{\sqrt{2}} \begin{pmatrix} \phi_2 + i \phi_1 \cr \phi_4 - i\phi_3 \end{pmatrix} .
\label{4dHiggs}
\end{align}
Without loss of generality, we assume
$\la \phi_1 \ra , \la \phi_2 \ra , \la \phi_3 \ra  =0$ and  
$\la \phi_4 \ra \not= 0$, 
which is related to the AB phase $\theta_H$ in the fifth dimension by
$\la \phi_4 \ra  = \theta_H f_H$, where
$f_H  = ({2}/{g_w}) \sqrt{ {k}/{L(z_L^2 -1)}}$.

The gauge symmetry breaking pattern of
$SU(3)_C\times SO(5)\times U(1)_X$ is given as 
\begin{align}
&SU(3)_C\times SO(5)\times U(1)_X \cr
\noalign{\kern 5pt}
&\hskip 0.5cm
\underset{BC}{\to} ~
SU(3)_C\times SU(2)_L\times SU(2)_R\times U(1)_X
 \ \ \mbox{at $y=0,L$}  \cr
\noalign{\kern 5pt}
&\hskip 0.5cm
\underset{\langle\Phi\rangle}{\to} ~
SU(3)_C\times SU(2)_L\times U(1)_Y
 \ \ \mbox{by\ the VEV $\langle\Phi_{({\bf 1},{\bf 4})}\rangle\not=0$  at\ $y=0$} \cr
\noalign{\kern 5pt}
&\hskip 0.5cm
\underset{\theta_H}{\to} ~
SU(3)_C\times U(1)_{EM}\ \ \ \  \mbox{by\ the\ Hosotani\ mechanism.}
\end{align}

\section{Observables}
\label{Sec:Observables}

\begin{figure}[tbh]
\begin{center}
\includegraphics[bb=0 0 426 90,height=2.2cm]{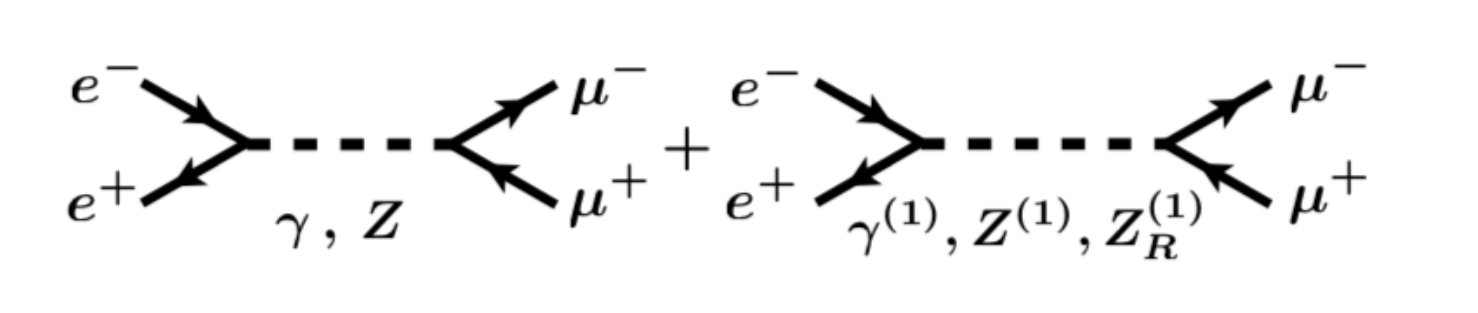}
\caption{{\small
 $e^-e^+\to \mu^-\mu^+$ process mediated by neutral vector bosons $V_i$.
 In the SM, $V_i=\gamma, Z$.
 In the GHU, $V_i=\gamma,Z,\gamma^{(1)},Z^{(1)},Z_R^{(1)}$.}
}
\label{Tab:ee-to-mumu}
\end{center}
\end{figure}

We quickly check observables in
$e^-e^+\to \{V_i\}\to \mu^-\mu^+$ shown in
Figure~\ref{Tab:ee-to-mumu}.
Observables in $e^-e^+\to \mu^-\mu^+$ process are summarized in 
Table~\ref{Tab:Observables-polarization}.
There are two observables (cross section $\sigma^{\mu^-\mu^+}$,
forward-backward (FB) asymmetry $A_{FB}^{\mu^-\mu^+}$)
in $e^-e^+\to \mu^-\mu^+$ process
for polarized initial state $e^-$ and $e^+$,
while there are four observables
(cross section $\sigma^{\mu^-\mu^+}$,
forward-backward (FB) $A_{FB}^{\mu^-\mu^+}$,
left-right (LR) $A_{LR}^{\mu^-\mu^+}$,
LR FB asymmetries $A_{LR,FB}^{\mu^-\mu^+}$).
This is because observable LR and LR FB asymmetries are zero for
unpolarized initial $e^-$ and $e^+$ states.
(See e.g., 
Refs.~\cite{Blondel:1987gp,Kennedy:1988rt,Abe:1994bj,Abe:1994bm,Abe:1995yh,MoortgatPick:2005cw}.)

\begin{table}[tbh]
\begin{center}
\begin{tabular}{|ll|cc|c}
\hline
 \rowcolor[gray]{0.9}
 \rowcolor[gray]{0.9}
 Observables&(Symbol)&Unpolarized&Polarized\\\hline
 Cross sections  &($\sigma^{\mu^-\mu^+}$)& Yes& Yes\\
 Forward-backward (FB) asymmetries  &($A_{FB}^{\mu^-\mu^+}$)& Yes& Yes\\
 Left-right (LR) asymmetries   &($A_{LR}^{\mu^-\mu^+}$)& No& Yes\\
 LR FB asymmetries&($A_{LR,FB}^{\mu^-\mu^+}$)& No& Yes\\
 \hline
\end{tabular}
\caption{{\small
 Observables in $e^-e^+\to \mu^-\mu^+$ process for unpolarized and
 polarized initial state $e^-$ and $e^+$.
 Yes/No stand for mesurable/unmesurable.}
}
\label{Tab:Observables-polarization}
\end{center}
\end{table}

The observables in $e^-e^+\to {\mu^-\mu^+}$, which are summarized in
Table~\ref{Tab:observalbes}, can be expressed by four amplitudes
$Q_{e_X\mu_Y}$ $(X,Y=L,R)$ and an effective electron polarization
$P_{\rm eff}$:
\begin{align}
Q_{e_{X} \mu_{Y}}
=
\sum_{i} \frac{g_{V_{i}e}^{X} g_{V_{i}\mu}^{Y}}{(s-m_{V_{i}}^{2}) + i
 m_{V_{i}}\Gamma_{V_{i}}},
\label{Eq:Qs-1} 
\end{align}
where  
$s$ is the square of the center-of-mass energy, 
$g_{V_{i}f}^{L/R}$ $(f=e,\mu)$ are  couplings of the left- and
right-handed fermion $f$ to the vector boson $V_{i}$, and  $m_{V_{i}}$
and $\Gamma_{V_{i}}$ are the mass and total decay width of $V_{i}$;
the effective electron polarization is defined as
\begin{align}
P_{\rm eff}\equiv\frac{P_{e^-}-P_{e^+}}{1-P_{e^-}P_{e^+}}.
\end{align}
where $P_{e^\pm}$ denotes longitudinal polarization of $e^\pm$.
$P_{e^\pm} = +1$ corresponds to purely right-handed $e^\pm$.
(For more information, see e.g., Appendix~\ref{Sec:Observables-app}.)

\begin{table}[tbh]
\begin{center}
\begin{tabular}{|c|c|}
\hline
 \rowcolor[gray]{0.9}
 Obs.&Linear combination of amplitudes 
 \\\hline
 $\sigma^{\mu^-\mu^+}$ &$(1-P_{\rm eff})(
	 {|Q_{e_L\mu_L}|^2}
	 +{|Q_{e_L\mu_R}|^2}
	 )
     +(1+P_{\rm eff})
	 ({|Q_{e_R\mu_R}|^2}	 
	 +{|Q_{e_R\mu_L}|^2}
	 )$\\[1em]
 $A_{FB}^{\mu^-\mu^+}$    &$(1-P_{\rm eff})(
	 {|Q_{e_L\mu_L}|^2}
	 -{|Q_{e_L\mu_R}|^2}
	 )
     +(1+P_{\rm eff})(
	 {|Q_{e_R\mu_R}|^2}
	 -{|Q_{e_R\mu_L}|^2}
	 )$\\[1em]
 $A_{LR}^{\mu^-\mu^+}$           &$
	 {|Q_{e_L\mu_L}|^2}
	 +{|Q_{e_L\mu_R}|^2}
	 -{|Q_{e_R\mu_R}|^2}	 
	 -{|Q_{e_R\mu_L}|^2}
	 $\\[1em]
 $A_{LR,FB}^{\mu^-\mu^+}$&$
	 {|Q_{e_L\mu_L}|^2}
	 -{|Q_{e_L\mu_R}|^2}
	 -{|Q_{e_R\mu_R}|^2}	 
	 +{|Q_{e_R\mu_L}|^2}
	 $\\
 \hline
\end{tabular}
 \caption{{\small
 Observables ($\sigma^{\mu^-\mu^+}$, $A_{FB}^{\mu^-\mu^+}$,
 $A_{LR}^{\mu^-\mu^+}$, $A_{LR,FB}^{\mu^-\mu^+}$)  can be
 expressed by linear combination of amplitudes, where overall factors
 are omitted and the final state muon mass is neglected. }
}
\label{Tab:observalbes}
\end{center}
\end{table}

In the following, we will show the results of $e^-e^+\to \mu^-\mu^+$
process with polarized $e^-e^+$, 
$(P_{e^-},P_{e^{+}})=(\mp0.8,\pm0.3)$,
at $\sqrt{s}=250$ GeV with
250 fb${}^{-1}$ data 
in the GHU B-model.
(Note that the ILC designs 80\% polarization of the electron beam and
30\% polarization of the positron beam according to the ILC Technical
Design Report
\cite{Behnke:2013xla,Baer:2013cma,Adolphsen:2013jya,Adolphsen:2013kya,Behnke:2013lya}.)

In the SM, $\gamma$ and $Z$ boson contribute to the
$e^-e^+\to \mu^-\mu^+$ process. 
In GHU $Z'$ bosons, $\gamma^{(n)}$, $Z^{(n)}$ and $Z_R^{(n)}$
($n \ge 1$) give additional 
contributions.
The masses of neutral higher KK vector bosons $Z^{(2k-1)}$, $Z^{(2k)}$,
$Z_R^{(k)}$, and $\gamma^{(k)}$ $(k\geq 1)$ almost linearly
increase as $k$. The couplings constants of them to left- and
right-handed electrons are decreasing when $k$ is increasing.
The contribution for the low-energy observables from 
each higher KK vector boson $Z^{(k)}$, $Z_R^{(k)}$, $\gamma^{(k)}$
$(k\geq 2)$ is sub-dominant.
In the following, we consider contributions for the low-energy
observables only from the first KK bosons $Z^{(1)}$, $Z_R^{(1)}$,
and $\gamma^{(1)}$.

In this talk, we focus on muon pair production, so we only list the
formulas in Appendix \ref{Sec:Observables-app}, which are available
when the center-of-mass energy $\sqrt{s}$ is much larger
than the mass of the final state fermion $m_f$.
We omit the non-zero final state fermion mass $m_f\not=0$ and each
statistical error estimation here. 
(For more information, see Ref.~\cite{Funatsu:2020haj}.)

\section{Parameter sets}
\label{Sec:Setup}

Before we calculate observables in the B-model, we need to specify
parameter sets in the model. In the following, we roughly describe how to
determine the parameter sets. For more information, see
Ref.~\cite{Funatsu:2020haj} and also 
Refs.~\cite{Funatsu:2019xwr,Funatsu:2019fry,Funatsu:2020znj}. 

Parameters of the model are determined in the following steps:
\begin{itemize}
 \item[(i)] We pick the values of $\theta_H$ and
	    $m_{\KK}=  \pi k (z_L -1)^{-1}$.
 \item[(ii)] $k$ is determined by reproducing the $Z$ boson mass $m_Z$.
	     The warped factor $z_L$ is also fixed by using
	     $m_{\KK}=  \pi k (z_L -1)^{-1}$.
 \item[(iii)] The bare Weinberg angle $\theta_W^0$ in
	      Eq.~(\ref{Eq:gY-sW}) with given $\theta_H$ 
	      is determined to fit the observed 
	      FB asymmetry
	      $A_{FB}^{\mu^-\mu^+}=0.0169\pm0.0013$
	      at $\sqrt{s}=m_Z$\cite{ALEPH:2005ema,Tanabashi:2018oca}.
\item[(iv)] By using the value of $\sin \theta_W^0$, wave functions of
	    gauge bosons are fixed.
\item[(v)]  The bulk mass parameters of
	    quark and lepton multiplets
	    $\Psi_{({\bf 3,4})}^\alpha$ and 
	    $\Psi_{({\bf 1,4})}^\alpha$ are fixed from the masses of
	    up-type quarks and charged leptons. 
\item[(vi)] The bulk mass parameters of down-type quark multiplets
	     $\Psi_{({\bf 3,1})}^{\pm \alpha}$ and brane interaction
	     coefficients in the down-quark sector are determined by 
	     reproducing the masses of down-type quarks.
	     The Majorana mass 
	    terms and brane interactions in the neutrino sector are
	     determined by reproducing neutrino masses.
\end{itemize}
The above processes must be repeated to satisfy all the  requirements
for each parameter set. By using the consistent parameter sets, we
proceed the following steps.
\begin{itemize}
\item[(a)] With these parameters fixed, wave functions of quarks and
	   leptons are determined. 
\item[(b)] The  $Z'$ coupling constants to the SM fermions, etc. are
	   determined.  
\item[(c)] The four-dimensional $Z'$ couplings of quarks and
	   leptons are obtained from the 5D gauge
	   interaction terms by inserting wave functions of gauge bosons
	   and quarks or leptons  and integrating over the
	   fifth dimensional coordinate
	   \cite{Funatsu:2014fda,Funatsu:2015xba,Funatsu:2016uvi}.  
\item[(d)] By using masses and couplings of $Z'$ bosons, decay
	   widths of $Z'$ bosons are calculated.
\end{itemize}
From the above procedures, we obtain some consistent parameter sets.
\begin{itemize}
\item The masses and widths of $\gamma$, $Z$ boson, and the first
      neutral KK vector bosons $Z^{(1)}$, $Z_R^{(1)}$, $\gamma^{(1)}$ are
      listed in  Table~\ref{Table:Mass-Width-Vector-Bosons}.
\item The coupling constants of $Z$ boson and the first neutral KK vector
      bosons $Z^{(1)}$, $Z_R^{(1)}$, $\gamma^{(1)}$ to quarks and leptons 
      are listed in Tables~\ref{Table:Couplings-Zprime_thetaH=010-mKK=13},
      \ref{Table:Couplings-Zprime_thetaH=010-mKK=11}, and
      \ref{Table:Couplings-Zprime_thetaH=010-mKK=15}.
\item Note that  for $\theta_H=0.10$ the top quark mass can be
      reproduced only 
      if $z_L\geq 10^{8.1}$ and dynamical electroweak symmetry breaking
      is achieved only if $z_L\leq 10^{15.5}$, the values of which
      correspond to $m_{\rm KK}\simeq[11,15] \,$TeV discussed in
      Ref.~\cite{Funatsu:2020znj}. 
\end{itemize}

\begin{table}[thb]
{
 \footnotesize
\begin{center}
\begin{tabular}{c|cc|cccccccc|c}
\hline
 \rowcolor[gray]{0.9}
 &&&&&&&&&&&\\[-0.75em]
\rowcolor[gray]{0.9}
 Name&$\theta_H$&$m_{\KK}$&$z_L$&$k$
 &$m_{\gamma^{(1)}}$&$\Gamma_{\gamma^{(1)}}$
 &$m_{Z^{(1)}}$&$\Gamma_{Z^{(1)}}$
 &$m_{Z_R^{(1)}}$&$\Gamma_{Z_R^{(1)}}$
 &Table\\
\rowcolor[gray]{0.9}
 &\mbox{[rad.]}&[TeV]&&[GeV]&[TeV]&[TeV]&[TeV]&[TeV]&[TeV]&[TeV]
 &\\ 
\hline 
 &&&&&&&&&&\\[-0.75em]
 \hspace{0.5em}B$^{\rm L}$&0.10&11.00&1.980$\times10^{8\ }$&6.933$\times10^{11}$&8.715&2.080&8.713&4.773&8.420&0.603
 &\ref{Table:Couplings-Zprime_thetaH=010-mKK=11}\\
\rowcolor[rgb]{0.55, 1.0, 1.0}
 B&0.10&13.00&3.865$\times10^{11}$ &1.599$\times10^{15}$&10.20&3.252&10.20&7.840&9.951&0.816
 &\ref{Table:Couplings-Zprime_thetaH=010-mKK=13}\\
 \hspace{0.5em}B$^{\rm H}$&0.10&15.00&2.667$\times10^{15}$&1.273$\times10^{19}$&11.69&4.885&11.69&11.82&11.48&1.253
 &\ref{Table:Couplings-Zprime_thetaH=010-mKK=15}\\
\hline
\end{tabular}\\[0.5em]
 \caption{{\small
 Masses and widths of $Z'$ bosons ($Z^{(1)}$, $\gamma^{(1)}$, and
 $Z_R^{(1)}$) are listed for
 $\theta_H=0.10$ and three $m_{\rm KK}=11,13,15\,$TeV values.
 $m_{Z}=91.1876\,$GeV and $\Gamma_{Z}=2.4952
 \,$GeV\cite{Tanabashi:2018oca}. 
 The column ``Name''   denotes each parameter set  
 and the column  ``Table'' indicate the table summarizing  coupling
 constants in each set.
 This is a part of Table III in Ref.~\cite{Funatsu:2020haj}.}
 }
\label{Table:Mass-Width-Vector-Bosons}
\end{center}
}
\end{table}

\begin{table}[htb]
{\small
 \begin{center}
\begin{tabular}{ccccccccc}
\hline
 \rowcolor[gray]{0.9}
 &&&&&&&&\\[-0.75em]
\rowcolor[gray]{0.9}
 $f$
 &$g_{Zf}^L$&$g_{Zf}^R$
 &$g_{Z^{(1)}f}^L$&$g_{Z^{(1)}f}^R$
 &$g_{Z_R^{(1)}f}^L$&$g_{Z_R^{(1)}f}^R$
 &$g_{\gamma^{(1)}f}^L$&$g_{\gamma^{(1)}f}^R$\\
 \hline
 \rowcolor[rgb]{0.55, 1.0, 1.0}
 $e$
 &$-$0.3058&\ \ 0.2629
 &$-$1.7621&$-$0.0584
 &$-$1.0444&0
 &$-$2.7587&\ \ 0.1071
 \\
 \rowcolor[rgb]{1, 1, 0.5}
 $\mu$
 &$-$0.3058&\ \ 0.2629
 &$-$1.6778&$-$0.0584
 &$-$0.9969&0
 &$-$2.6268&\ \ 0.1071
 \\
\hline
\end{tabular}
 \caption{{\small
 Coupling constants of neutral vector bosons, $Z'$ bosons, to
 fermions in units of $g_w=e/\sin\theta_W^0$
 are listed  for $\theta_H=0.10$ and $m_{\KK}=13.00 \,$TeV 
 (B) in Table~\ref{Table:Mass-Width-Vector-Bosons},
 where $\sin^2\theta_W^0=0.2306$. 
 Their corresponding $Z$ boson coupling constants in the SM are
 $(g_{Z_{e}}^L,g_{Z_{e}}^R)=(-0.3065,0.2638)$.
 Their corresponding $\gamma$ boson coupling constants are the same as
 those in the SM. When the value is less than $10^{-4}$, we write $0$.
 This is a part of Table IV in Ref.~\cite{Funatsu:2020haj}, which
 contains the other coupling constants.}
 }
\label{Table:Couplings-Zprime_thetaH=010-mKK=13}
 \end{center}
}
\end{table}

\begin{table}[htb]
 {\small
 \begin{center}
\begin{tabular}{ccccccccc}
\hline
 \rowcolor[gray]{0.9}
 &&&&&&&&\\[-0.75em]
\rowcolor[gray]{0.9}
 $f$
 &$g_{Zf}^L$&$g_{Zf}^R$
 &$g_{Z^{(1)}f}^L$&$g_{Z^{(1)}f}^R$
 &$g_{Z_R^{(1)}f}^L$&$g_{Z_R^{(1)}f}^R$
 &$g_{\gamma^{(1)}f}^L$&$g_{\gamma^{(1)}f}^R$\\
 \hline
 \rowcolor[rgb]{0.55, 1.0, 1.0}
 $e$
 &$-$0.3058&\ \ 0.2629
 &$-$1.5398&$-$0.0695
 &$-$0.9143&0
 &$-$2.4107&\ \ 0.1274
 \\
 \rowcolor[rgb]{1, 1, 0.5}
 $\mu$
 &$-$0.3058&\ \ 0.2629
 &$-$1.4545&$-$0.0695
 &$-$0.8670&0
 &$-$2.2772&\ \ 0.1274
 \\
\hline
\end{tabular}
 \caption{{\small
 Coupling constants of neutral vector bosons, $Z'$ bosons, to
 fermions in units of $g_w=e/\sin\theta_W^0$
 are listed  for
 $\theta_H=0.10$ and $m_{\KK}=11.00 \,$TeV
 (B$^{\rm L}$) in  Table~\ref{Table:Mass-Width-Vector-Bosons},
 where $\sin^2\theta_W^0=0.2306$.
Other information is the same as in
 Table~\ref{Table:Couplings-Zprime_thetaH=010-mKK=13}.
 This is a part of Table V in Ref.~\cite{Funatsu:2020haj}.}
 }
\label{Table:Couplings-Zprime_thetaH=010-mKK=11}
 \end{center}
}
\end{table}

\begin{table}[htb]
 {\small
 \begin{center}
\begin{tabular}{ccccccccc}
\hline
 \rowcolor[gray]{0.9}
 &&&&&&&&\\[-0.75em]
\rowcolor[gray]{0.9}
 $f$
 &$g_{Zf}^L$&$g_{Zf}^R$
 &$g_{Z^{(1)}f}^L$&$g_{Z^{(1)}f}^R$
 &$g_{Z_R^{(1)}f}^L$&$g_{Z_R^{(1)}f}^R$
 &$g_{\gamma^{(1)}f}^L$&$g_{\gamma^{(1)}f}^R$\\
 \hline
 \rowcolor[rgb]{0.55, 1.0, 1.0}
 $e$
 &$-$0.3057&\ \ 0.2629
 &$-$1.9841&$-$0.0504
 &$-$1.1740&0
 &$-$3.1063&\ \ 0.0924
 \\
 \rowcolor[rgb]{1, 1, 0.5}
 $\mu$
 &$-$0.3057&\ \ 0.2629
 &$-$1.9033&$-$0.0504
 &$-$1.1279&0
 &$-$2.9780&\ \ 0.0924
 \\
\hline
\end{tabular}
 \caption{{\small
 Coupling constants of neutral vector bosons, $Z'$ bosons, to
 fermions in units of $g_w=e/\sin\theta_W^0$
 are listed for  $\theta_H=0.10$ and $m_{\KK}=15.00\,$TeV
 (B$^{\rm H}$) in Table~\ref{Table:Mass-Width-Vector-Bosons},
 where $\sin^2\theta_W^0=0.2306$. 
Other information is the same as in
 Table~\ref{Table:Couplings-Zprime_thetaH=010-mKK=13}.
 This is a part of Table VI in Ref.~\cite{Funatsu:2020haj}.}
 }
\label{Table:Couplings-Zprime_thetaH=010-mKK=15}
 \end{center}
 }
\end{table}

From Tables~\ref{Table:Couplings-Zprime_thetaH=010-mKK=13},
\ref{Table:Couplings-Zprime_thetaH=010-mKK=11},
\ref{Table:Couplings-Zprime_thetaH=010-mKK=15},
we find that the coupling constants of the first neutral KK vector 
bosons $Z^{(1)}$, $Z_R^{(1)}$, $\gamma^{(1)}$ to quarks and leptons
are larger than those of the right-handed fermions 
except for $Z_R^{(1)}$ couplings to the top and bottom quarks.

\section{Results}
\label{Sec:Reults}

\begin{figure}[thb]
\begin{center}
\includegraphics[bb=0 0 504 344,height=5cm]{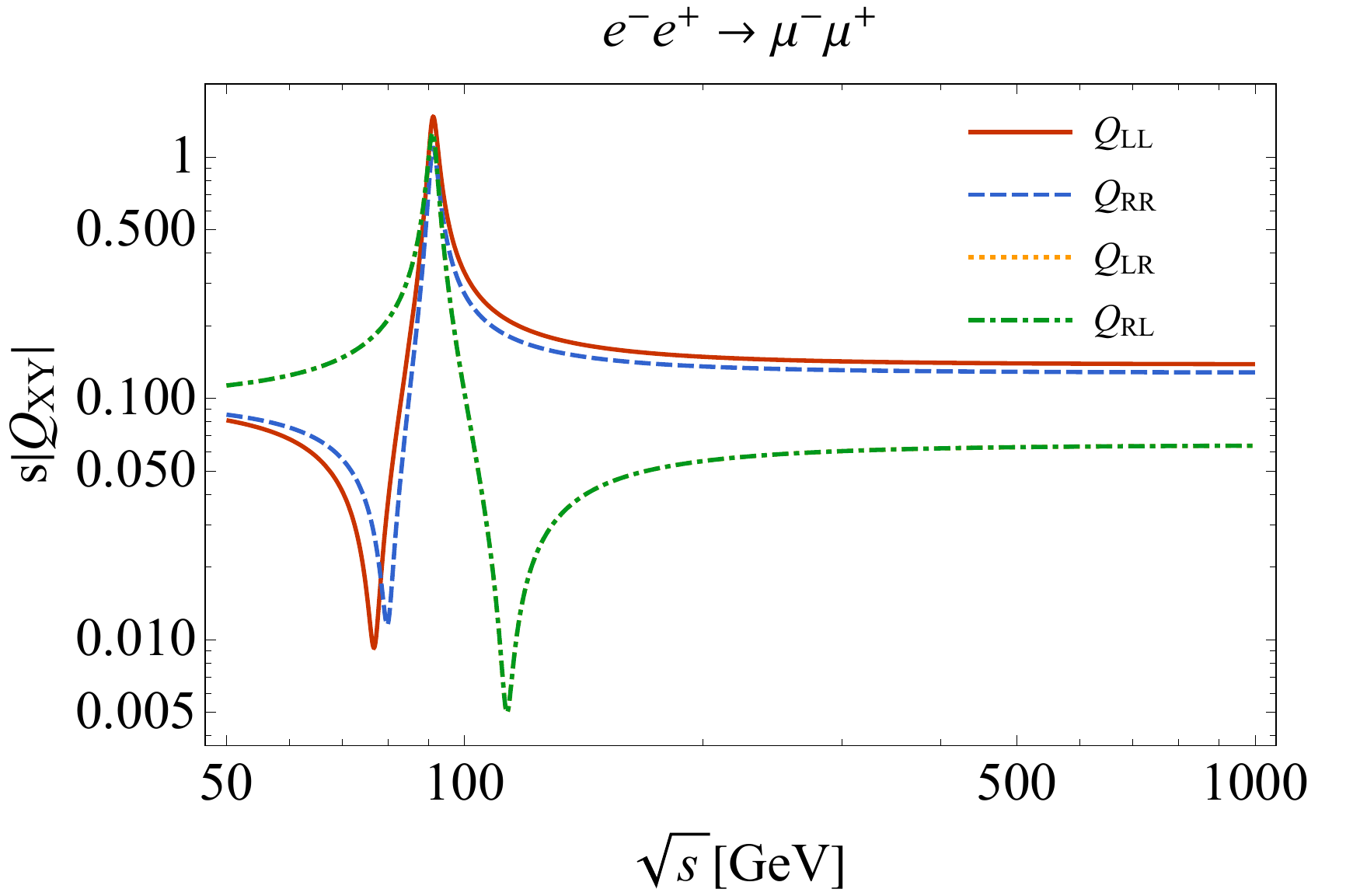}
\includegraphics[bb=0 0 504 338,height=5cm]{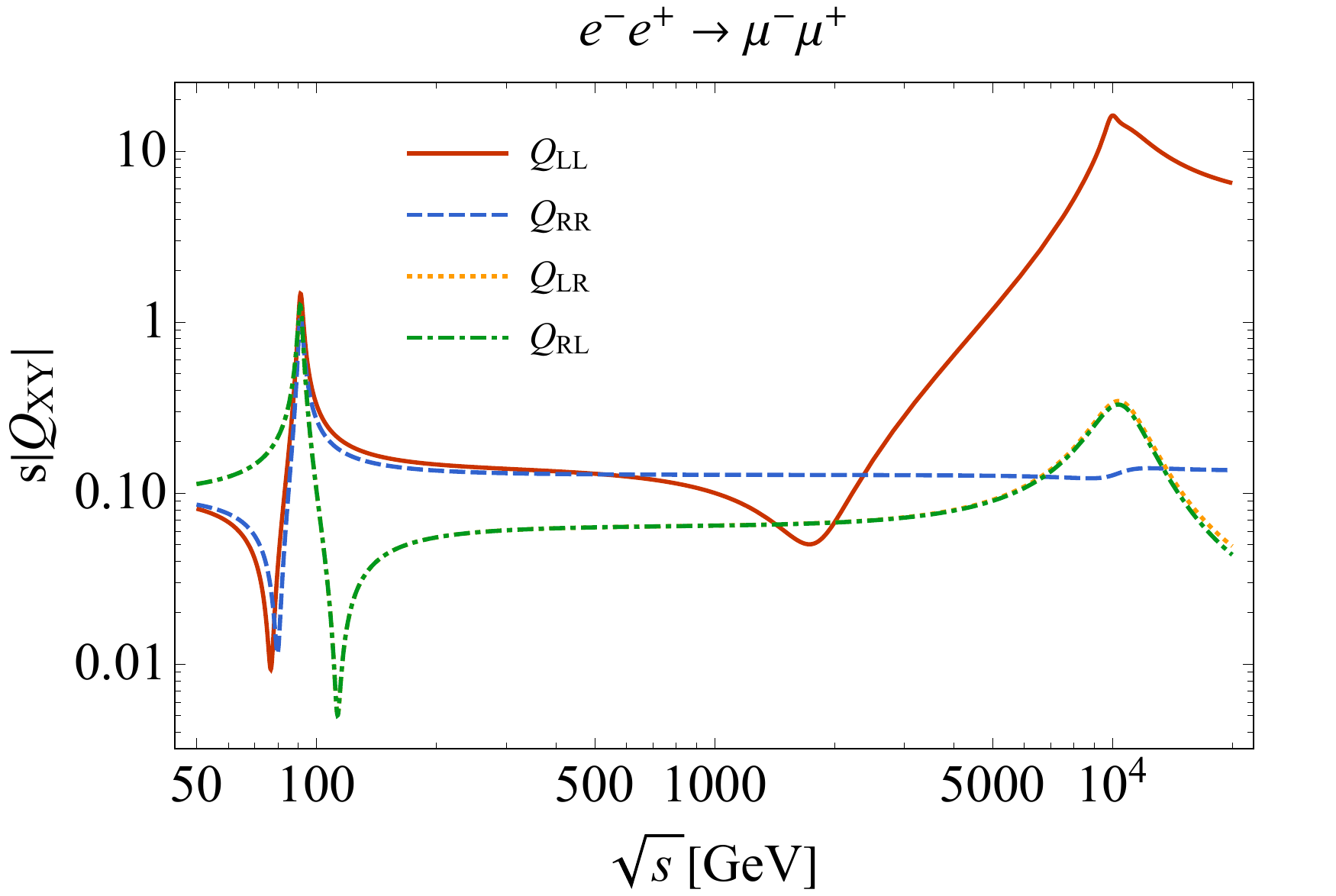}\\
 \caption{{\small
 The amplitude $s|Q_{e_X \mu_Y}|$
 $(X,Y=L, R)$ vs $\sqrt{s}\,$[GeV] for
 the SM (left figure)
 and  the GHU (B) (right figure) in
 Table~\ref{Table:Mass-Width-Vector-Bosons} 
 are shown.
 In each figure $Q_{e_X \mu_Y}$ is denoted as $Q_{X Y}$.
 The energy ranges $\sqrt{s}$ in the left and right figures are
 $\sqrt{s}=[50,1000] \,$GeV and $[50,2\times 10^4] \,$GeV, respectively.
 The figures come from Figure~4 in Ref.~\cite{Funatsu:2020haj}.}
 } 
 \label{Figure:sQ_ef-s_dependence}
\end{center}
\end{figure}

\begin{figure}[thb]
\begin{center}
\includegraphics[bb=0 0 504 341,height=5cm]{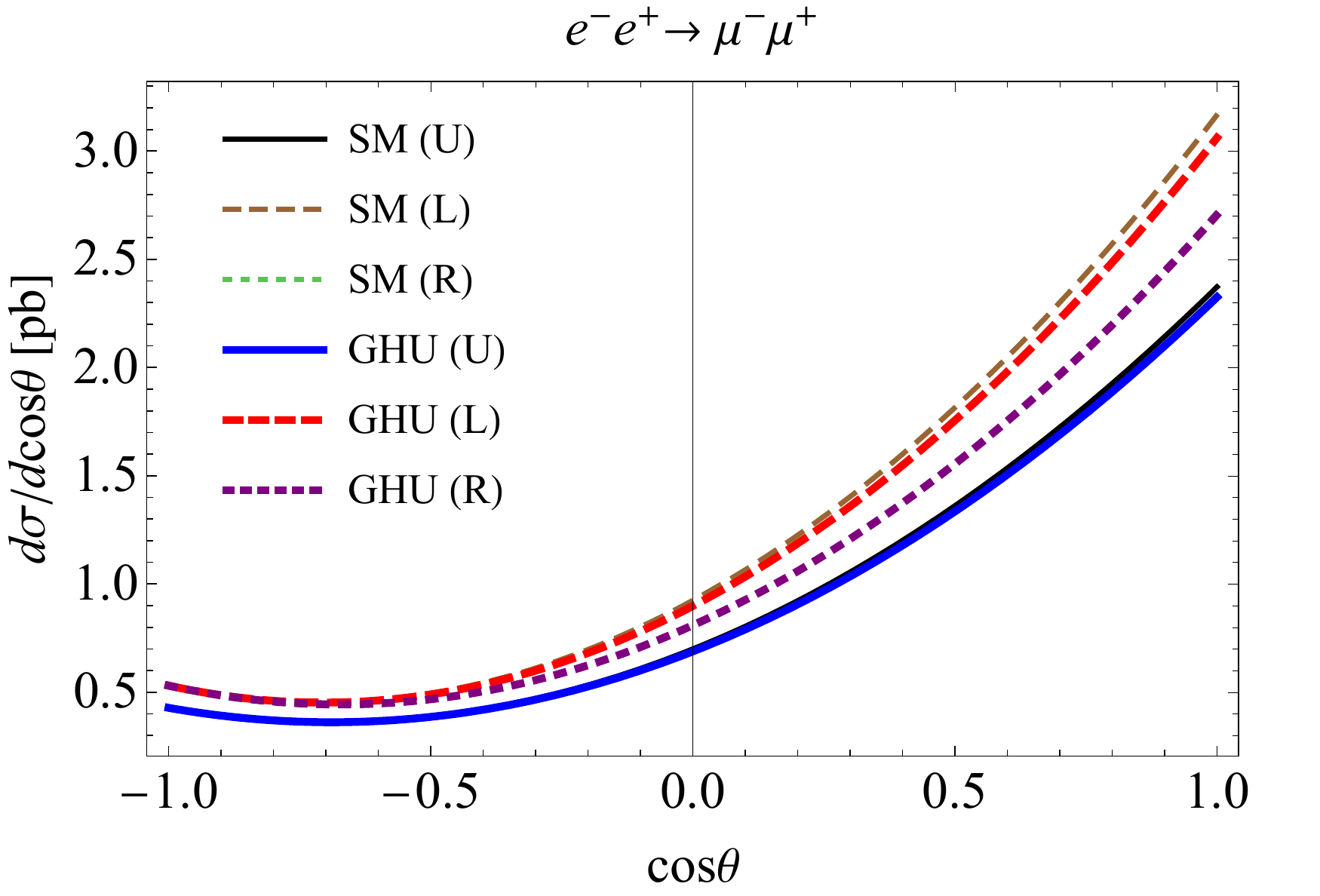}
\includegraphics[bb=0 0 504 327,height=5.cm]{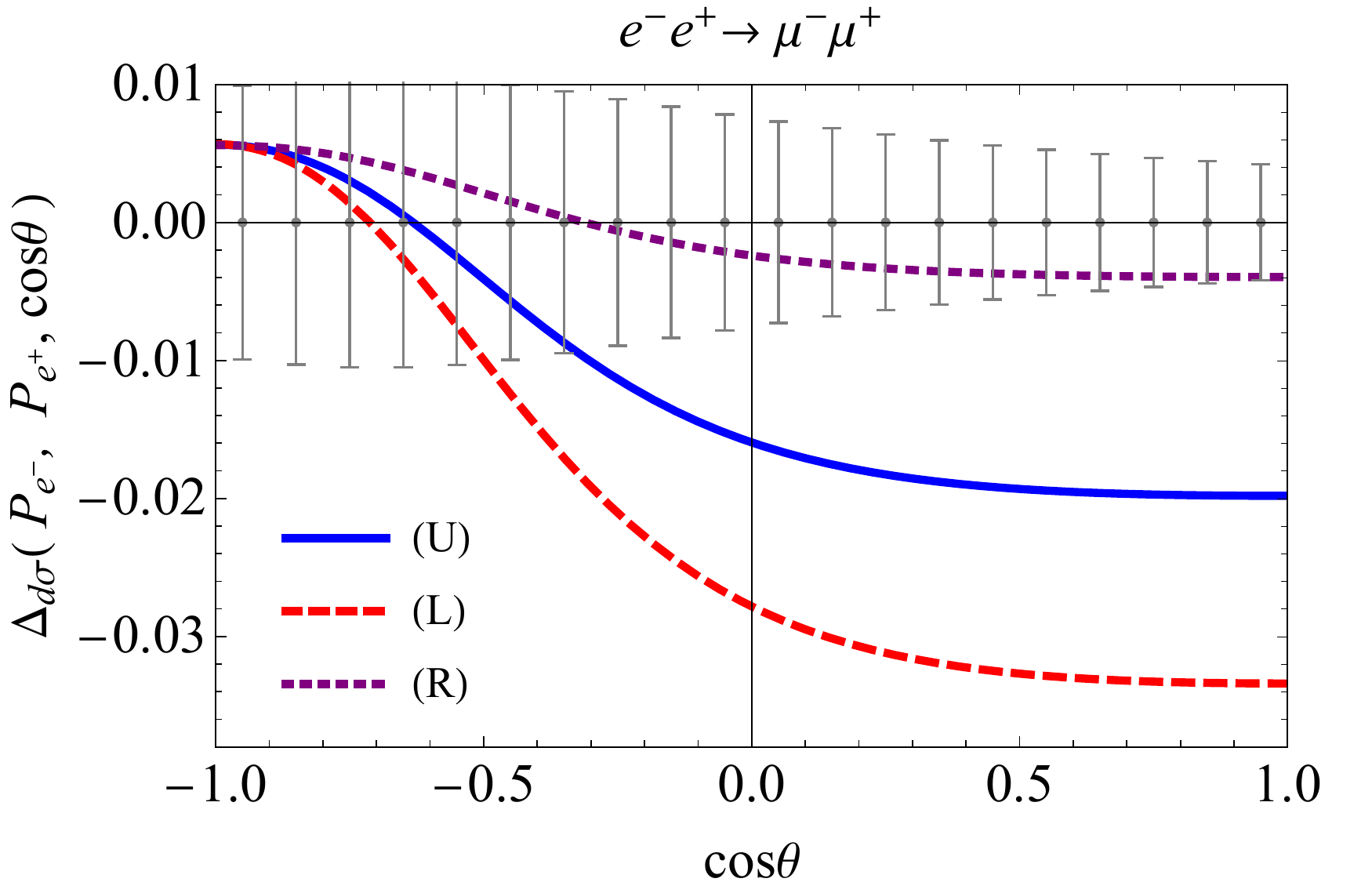}\\
 \caption{{\small
 Differential cross section  $d\sigma^{\mu^-\mu^+}/d\cos\theta$
 is shown.
 The left figure shows the $\theta$ dependence of
 $d\sigma^{\mu^-\mu^+}/d\cos\theta$
in the SM and the GHU (B) in  Table~\ref{Table:Mass-Width-Vector-Bosons}
 with three sets
 $(P_{e^-},P_{e^+})=(0,0) (U), (-0.8,+0.3) (L),  (+0.8,-0.3) (R)$.  
 $\sqrt{s}=250 \,$GeV.
 The right figure shows the $\theta$ dependence of 
 $\Delta_{d\sigma}^{f\bar{f}}(P_{e^-},P_{e^+}, \cos\theta )$ in
 Eq.~(\ref{Eq:Delta_dsigma}). 
 The error bars represent statistical errors in the SM 
 at $\sqrt{s}=250 \,$GeV with 250$\,$fb$^{-1}$ data.
 Each bin is given by $\cos\theta=[k-0.05,k+0.05]$
 ($k=-0.95,-0.85,\cdots,0.95$).
 The figures come from Figure~5 in Ref.~\cite{Funatsu:2020haj}.}
 } 
 \label{Figure:dsigma-ef-theta=010}
\end{center}
\end{figure}

\begin{figure}[thb]
\begin{center}
\includegraphics[bb=0 0 504 336,height=5cm]{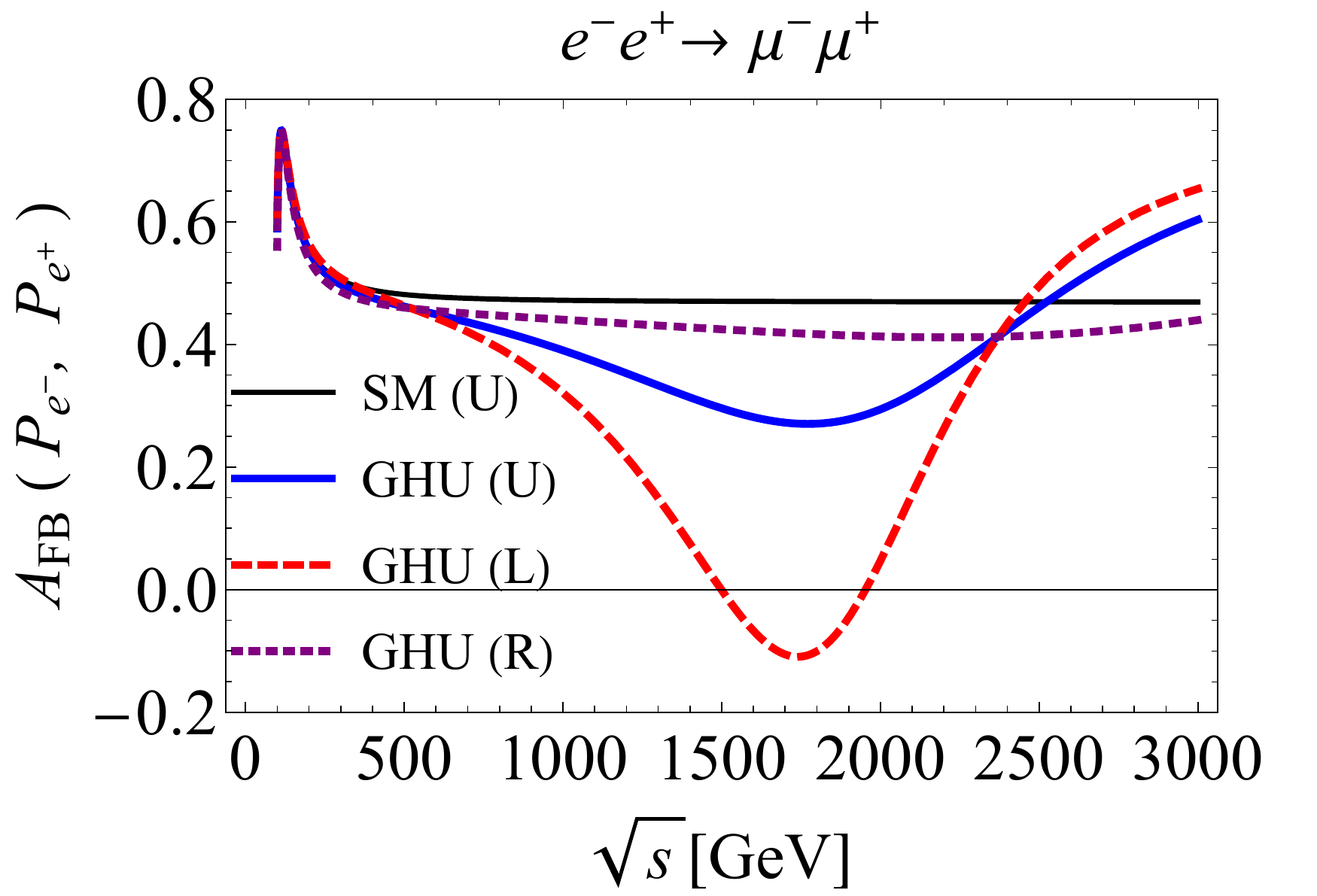}
\includegraphics[bb=0 0 504 320,height=5cm]{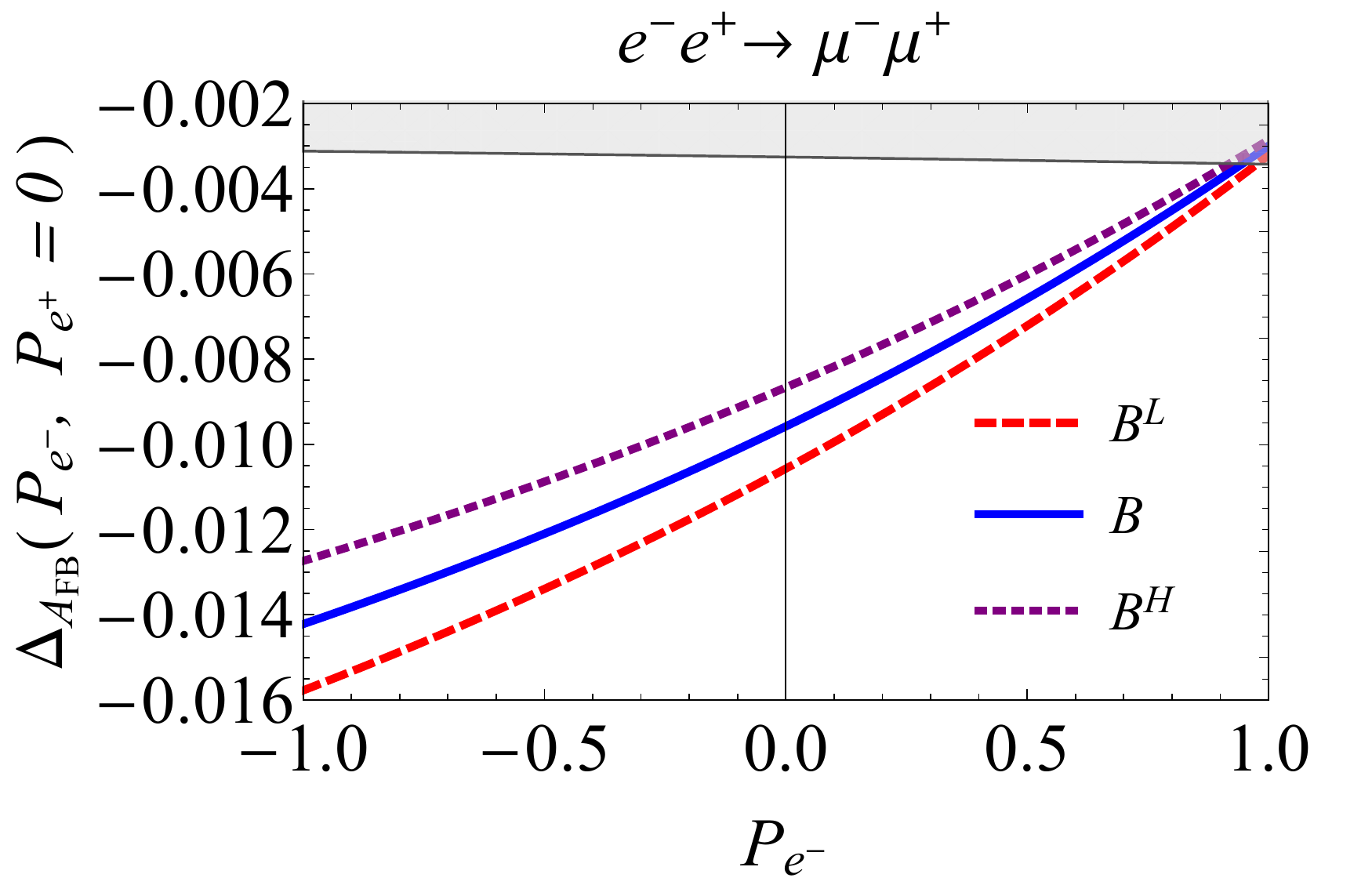}
 \caption{{\small
 Forward-backward asymmetry $A_{FB}^{\mu^-\mu^+}$
 is shown.
 The left figure shows the $\sqrt{s}$ dependence of
 $A_{FB}^{f\bar{f}}$  in the SM and the GHU  (B) in
 Table~\ref{Table:Mass-Width-Vector-Bosons}. 
 Three cases of polarization of electron  and positron beams 
 $(P_{e^-},P_{e^+})=(0,0)(U), (-0.8,+0.3) (L), (+0.8,-0.3)(R)$ are
 depicted for GHU. 
 The energy range $\sqrt{s}$ is  $[80,3000] \,$GeV.
 The right figure shows the electron polarization $P_{e^-}$
 dependence of the deviation from the SM
 $\Delta_{A_{FB}}^{f\bar{f}}(P_{e^-},P_{e^+}=0)$ in
 Eq.~(\ref{Eq:Delta_A_FB}) 
 for the GHU  (B$^{\rm L}$), (B), (B$^{\rm H}$) in
 Table~\ref{Table:Mass-Width-Vector-Bosons}. 
 The gray band in the central and right side figures represent
 the statistical error in the SM  at $\sqrt{s}=250 \,$GeV with
 250$\,$fb$^{-1}$ data.
 The figures come from Figure~6 in Ref.~\cite{Funatsu:2020haj}.}
 } 
 \label{Figure:AFB-ef-theta=010}
\end{center}
\end{figure}

\begin{figure}[thb]
\begin{center}
\includegraphics[bb=0 0 504 332,height=5cm]{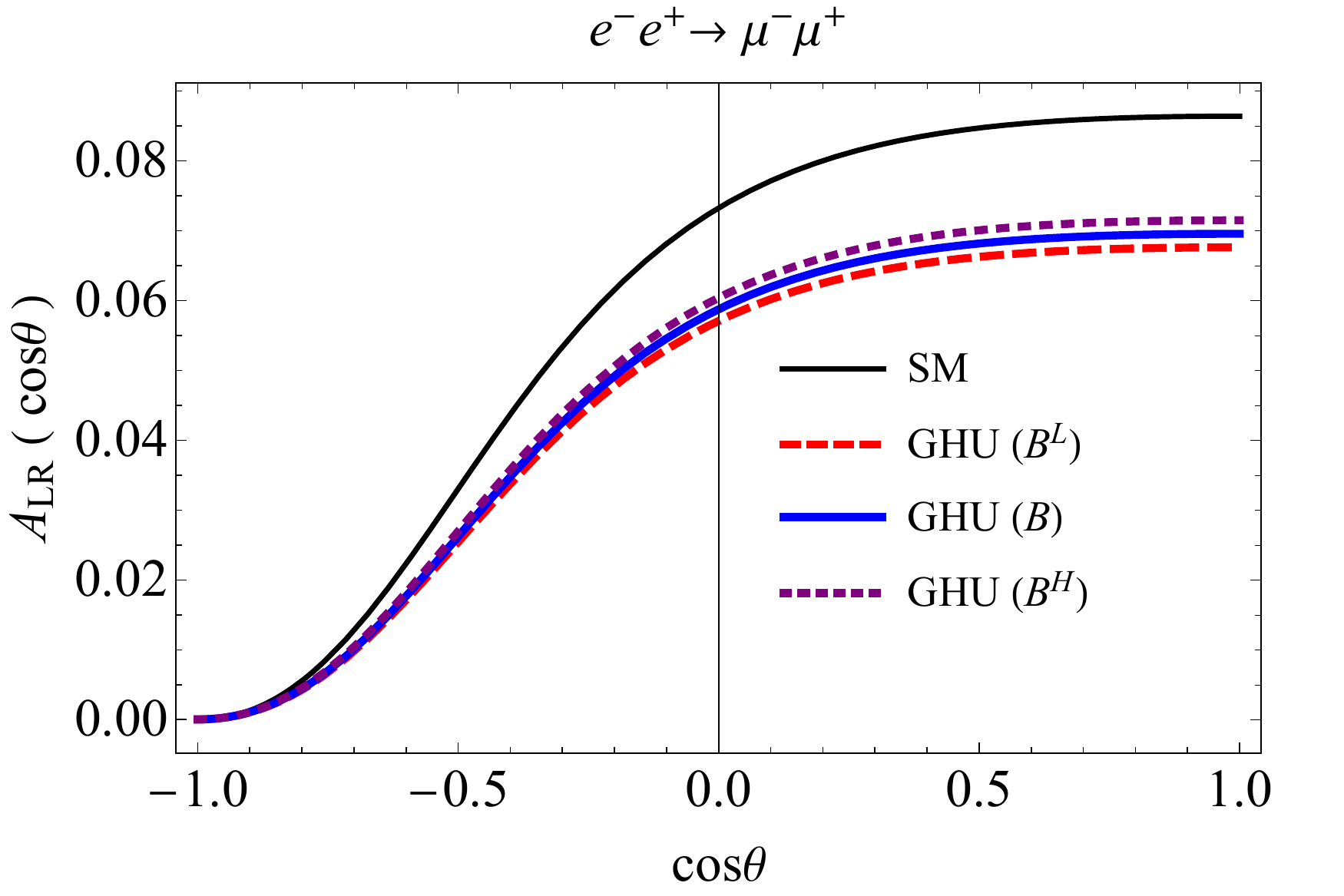}
\includegraphics[bb=0 0 504 332,height=5cm]{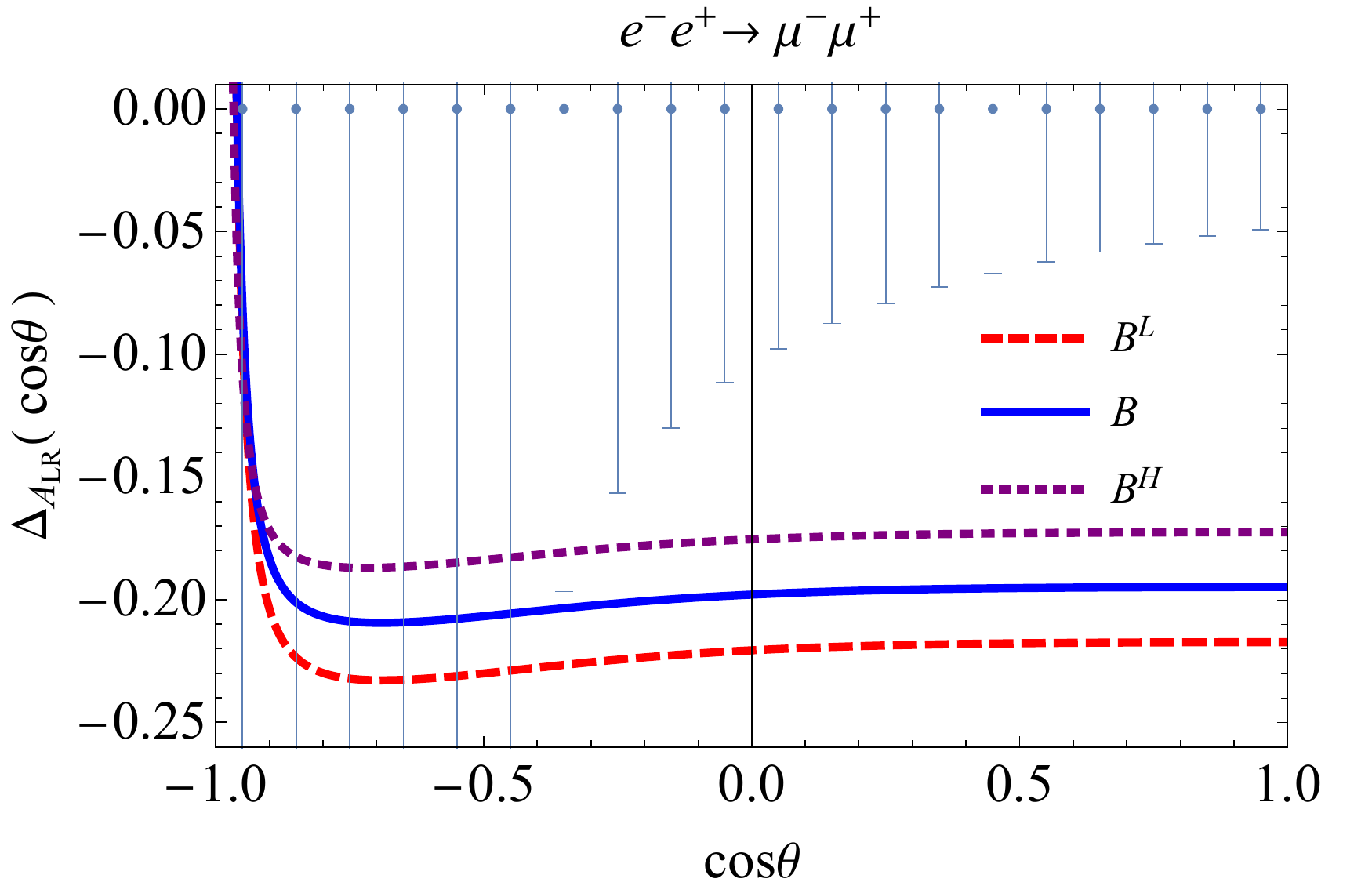}\\
 \caption{{\small
 Left-right asymmetry
 $A_{LR}^{\mu^-\mu^+}(\cos\theta)$
 is shown.
 The left figure shows the $\theta$ dependence of
 $A_{LR}^{\mu^-\mu^+}(\cos\theta)$ 
 for the SM and the GHU
 (B$^{\rm L}$), (B), (B$^{\rm H}$) in
 Table~\ref{Table:Mass-Width-Vector-Bosons}. 
 The right figure shows the $\theta$ dependence of
 the deviation of the left-right asymmetry from the SM, 
 $\Delta_{A_{LR}}^{\mu^-\mu^+}(\cos\theta)$
 in Eq.~(\ref{Eq:Delta_A_LR})  for  the GHU  (B$^{\rm L}$), (B),
 (B$^{\rm H}$). 
 The error bars in the right side figures represent
 the statistical error 
 at $\sqrt{s}=250 \,$GeV with 250$\,$fb$^{-1}$ data and
 $(P_{e^-},P_{e^+})=(-0.8,+0.3), (+0.8,-0.3)$.
 The figures come from Figure~8 in Ref.~\cite{Funatsu:2020haj}.}
 } 
 \label{Figure:ALR-cos-ef-theta=010}
\end{center}
\end{figure}

\begin{figure}[thb]
\begin{center}
\includegraphics[bb=0 0 504 327,height=5cm]{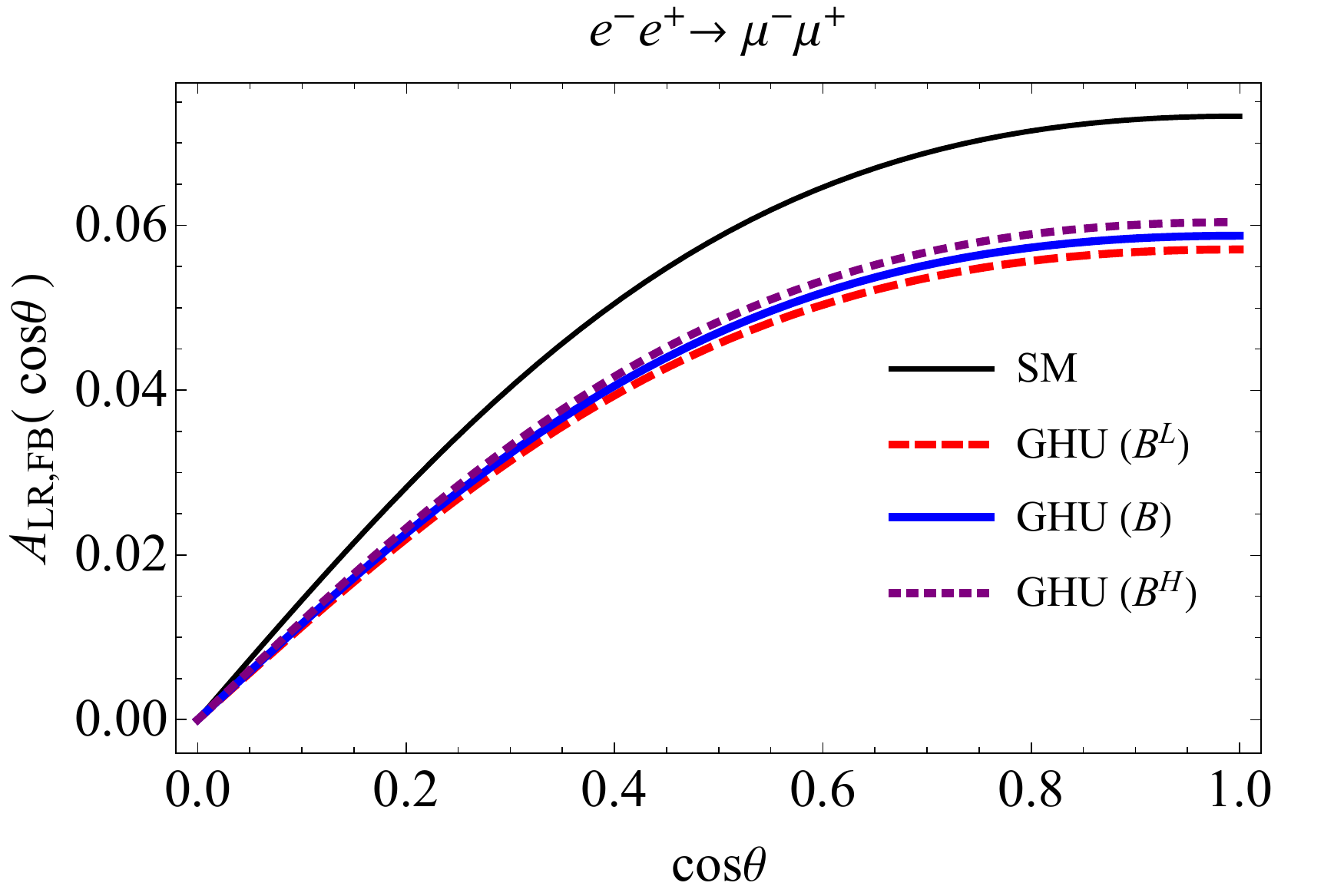}
\includegraphics[bb=0 0 504 325,height=5cm]{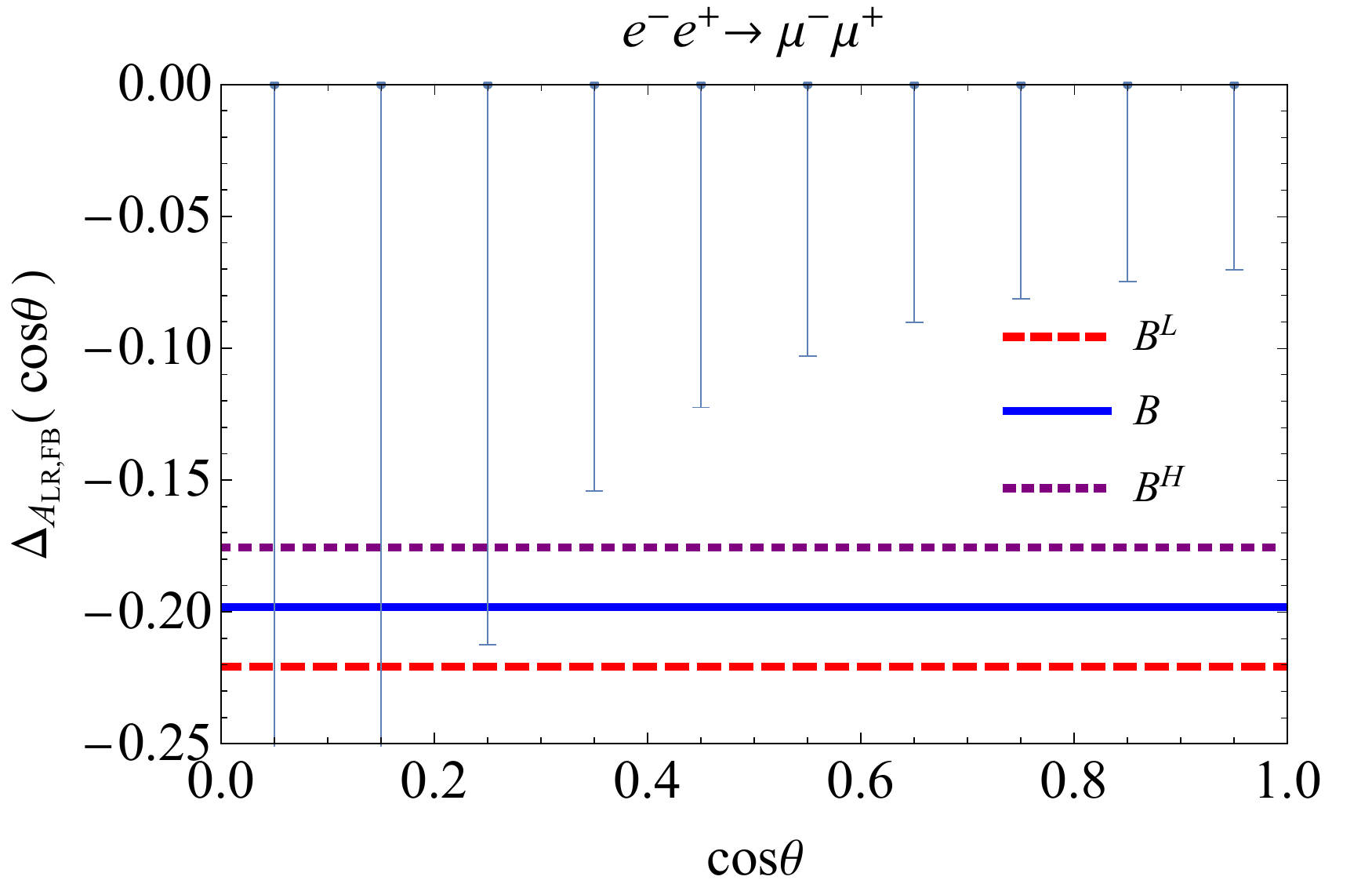}\\
 \caption{{\small
 Left-right forward-backward asymmetry
 $A_{LR,FB}^{\mu^-\mu^+}(\cos\theta)$
 is shown.
 The left figure shows the $\sqrt{s}$ dependence of
 $A_{LR,FB}^{\mu^-\mu^+}(\cos\theta)$ 
 for the SM and the GHU
 (B$^{\rm L}$), (B), (B$^{\rm H}$) in
 Table~\ref{Table:Mass-Width-Vector-Bosons}. 
 The right figure shows the $\cos\theta$  dependence of  the deviation
 of  the left-right asymmetry from the SM, 
 $\Delta_{A_{LR, FB}}(\cos\theta)$
 in Eq.~(\ref{Eq:Delta-A_LRFB}) for  the GHU (B$^{\rm L}$), (B),
 (B$^{\rm H}$) in 
 Table~\ref{Table:Mass-Width-Vector-Bosons}. 
 The error bars in the right side figures stand for
 the statistical error
 at $\sqrt{s}=250 \,$GeV with 250$\,$fb$^{-1}$ data and
 $(P_{e^-},P_{e^+})=(-0.8,+0.3), (+0.8,-0.3)$.
 The figures come from Figure~9 in Ref.~\cite{Funatsu:2020haj}.}
 } 
 \label{Figure:ALRFB-ef-theta=010}
\end{center}
\end{figure}

In this section, we show observables of the $s$-channel scattering
process of $e^-e^+\to \mu^-\mu^+$ mediated by  neutral vector bosons $V$
in the GHU, where $V =\gamma,Z,\gamma^{(1)},Z^{(1)},Z_R^{(1)}$.

\subsection{Amplitude}

Cross sections are determined in terms of amplitudes $Q_{e_{X} \mu_{Y}}$
($X, Y = L, R$) in Eq.~(\ref{Eq:Qs-1}). 
In Figure~\ref{Figure:sQ_ef-s_dependence}, $\sqrt{s}$-dependence of
$s |Q_{e_{X} \mu_{Y}}|$ is displayed for the SM and GHU (B).

In the SM,
\begin{align}
Q_{e_{X} \mu_{Y}}^\SM
= \frac{e^2}{s}  +\frac{g_{Ze}^Xg_{Z\mu}^Yg_{w}^2}{(s-m_{Z}^{2}) + i m_{Z}\Gamma_{Z}}.
\end{align}
$s |Q_{e_{X} \mu_{Y}}|$ has peak at $\sqrt{s}=m_Z$ and
$Q_{e_{L} \mu_{R}} = Q_{e_{R} \mu_{L}}$. 
$Q_{e_{L} \mu_{L}} =Q_{e_{R} \mu_{R}}$ becomes smaller below
$\sqrt{s}=m_Z$ and  $Q_{e_{L} \mu_{R}}$ and $Q_{e_{R} \mu_{L}}$ become
smaller above $\sqrt{s}=m_Z$ as a result of the interference of the
$\gamma$ and $Z$ amplitudes. For $\sqrt{s}\gg m_{Z}$,
$sQ_{e_X\mu_Y} \simeq e^2+g_{Ze}^Xg_{Z\mu}^Yg_w^2$.
Furthre, the sign of the left-handed copling constant $g_{Zf}^L$
$(f=e,\mu)$ is opposite to that of the right-handed coupling constant
$g_{Zf}^R$, so
$|Q_{e_L\mu_L}|,|Q_{e_R\mu_R}|\gg |Q_{e_L\mu_R}|,|Q_{e_R\mu_L}|$.

In GHU,
\begin{align}
Q_{e_{X} \mu_{Y}} = Q_{e_{X} \mu_{Y}}^\SM + Q_{e_{X} \mu_{Y}}^{Z'}, \ \
Q_{e_{X} \mu_{Y}}^{Z'}  \simeq \sum_{V = \gamma^{(1)},Z^{(1)}, Z_R^{(1)}}
 \frac{g_{V e}^X g_{V \mu}^Y g_{w}^2} {(s-m_V^2) + i m_V \Gamma_V}  ~,
\end{align}
where we retained contributions from first KK modes in
$Q_{e_{X} \mu_{Y}}^{Z'}$.
For $\sqrt{s} \lesssim 200\,$GeV,
$Q_{e_{X} \mu_{Y}} \sim Q_{e_{X} \mu_{Y}}^\SM $.
In Figure~\ref{Figure:sQ_ef-s_dependence} the $\sqrt{s}$-dependence of
$s |Q_{e_{X} \mu_{Y}}|$ is plotted.  $Q_{e_{X} \mu_{Y}}$ has a peak
around $\sqrt{s}\simeq m_{Z'}\simeq 10 \,$TeV. 
The dominant component is $Q_{e_{L} \mu_{L}}$, which develops
significant deviation from the SM. $Q_{e_{L} \mu_{L}}$ has a dip around
$\sqrt{s}\simeq 1.7 \,$TeV. 

\subsection{Cross section}

The differential cross section of $e^-e^+\to\mu^-\mu^+$
$d\sigma^{\mu^-\mu^+} /d\cos\theta$ is shown at $\sqrt{s}=250\,$GeV
in Figure~\ref{Figure:dsigma-ef-theta=010}.
The differential cross section in the forward region are larger than  
that of the backward region regardless of the
polarization. The deviation from the SM are seen in the forward region
with less statistical errors. The differential cross section of the
100\% left- and right-handed polarized initial electron is
given by the formulas in Eq.~(\ref{Eq:dsigma_LR-RL-mf=0}).

In the SM the $Z$ couplings are different for left-handed and
right-handed fermions which leads to
$Q_{e_{L} \mu_{L}} \not= Q_{e_{L} \mu_{R}}$ and 
$Q_{e_{R} \mu_{R}} \not= Q_{e_{R} \mu_{L}}$.

In GHU, coupling constants of the left-handed fermions to
$Z'$ bosons are much larger than those of the right-handed ones.
The magnitude of the left-handed fermion couplings is 
rather large so that the amount of the deviation in
$d\sigma^{\mu^-\mu^+}/d\cos\theta$ from the SM becomes large for 
left-handed polarized electron beams, whereas the
deviation becomes small for right-handed electron beams.
$\Delta_{d\sigma}^{\mu^-\mu^+}(P_{e^-},P_{e^+}, \cos \theta)$ in
Eq.~(\ref{Eq:Delta_dsigma}) 
is plotted in the right figure of Figure~\ref{Figure:dsigma-ef-theta=010}.
The deviation for the parameter set (B) in
Table~\ref{Table:Mass-Width-Vector-Bosons} can be clearly seen in
$e^- e^+$ collisions at $\sqrt{s}=250 \,$GeV  
with 250$\,$fb$^{-1}$ data.

\subsection{Forward-backward asymmetry}

The FB asymmetry $A_{FB}^{\mu^-\mu^+}$ is shown in
Figure~\ref{Figure:AFB-ef-theta=010}. 
From Eq.~(\ref{Eq:A_FB}), 
$A_{FB}^{\mu^-\mu^+}(P_{e^-},P_{e^+})$ with
$(P_{e^-},P_{e^+})=(0,0),(-1,0),(+1,0)$ 
are given by 
\begin{align}
A_{FB}^{\mu^-\mu^+}(0, 0) ~ &\simeq
\frac{3}{4}\frac{
\{|Q_{e_{R} \mu_{R}}|^{2}+|Q_{e_{L} \mu_{L}}|^{2} \}
-\{|Q_{e_{R} \mu_{L}}|^{2} +|Q_{e_{L} \mu_{R}}|^{2} \}
}{
\{|Q_{e_{R} \mu_{R}}|^{2} +|Q_{e_{L} \mu_{L}}|^{2} \}
+\{|Q_{e_{R} \mu_{L}}|^{2} +|Q_{e_{L} \mu_{R}}|^{2} \} } ~, \cr 
A_{FB}^{\mu^-\mu^+}(-1, 0) &\simeq
 \frac{3}{4}\frac{|Q_{e_{L} \mu_{L}}|^{2}-|Q_{e_{L} \mu_{R}}|^{2}}
 {|Q_{e_{L} \mu_{L}}|^{2}+|Q_{e_{L} \mu_{R}}|^{2}} ~, \cr
A_{FB}^{\mu^-\mu^+}(1, 0) ~ &\simeq
 \frac{3}{4}\frac{|Q_{e_{R} \mu_{R}}|^{2}-|Q_{e_{R} \mu_{L}}|^{2}}
 {|Q_{e_{R} \mu_{R}}|^{2}+|Q_{e_{R} \mu_{L}}|^{2}}.
\end{align}

In the SM, the FB asymmetry $A_{FB}^{\mu^-\mu^+}$
becomes constant for $\sqrt{s}\gg m_Z$.
$A_{FB}^{\mu^-\mu^+}(P_{e^-},P_{e^+})$ becomes small at $Z$-pole
$\sqrt{s}=m_Z$. 
$A_{FB}^{\mu^-\mu^+}(P_{e^-},P_{e^+})\simeq 3/4$ slightly above $Z$-pole
$\sqrt{s}=m_Z$.
$A_{FB}^{\mu^-\mu^+}(P_{e^-},P_{e^+})$ approaches constant for
$\sqrt{s}\gg m_Z$. 

In the GHU (B) in Table~\ref{Table:Mass-Width-Vector-Bosons}, due to the
interference effects among $Z$ and $Z'$ bosons,
$|Q_{e_L\mu_L}|$ can be smaller than $|Q_{e_L\mu_R}|$ in some energy
region (around $\sqrt{s} \sim 1.7\,$TeV).
Consequently $A_{FB}^{\mu^-\mu^+}$ can become negative even for
$\sqrt{s}\gg m_Z$ 
as shown in Figure~\ref{Figure:AFB-ef-theta=010}.
Deviation from the SM starts to show up around $\sqrt{s} = 250\,$GeV.
As shown in the right figure in Figure~\ref{Figure:AFB-ef-theta=010},
the amount of the deviation $\Delta_{FB}^{\mu^-\mu^+}(P_{e^-},P_{e^+}=0)$
in Eq.~(\ref{Eq:Delta_A_FB}) becomes significant for $P_{e^-} \sim -1$
even at $\sqrt{s} = 250\,$GeV.

\subsection{Left-right asymmetry}

The LR asymmetry $A_{LR}^{\mu^-\mu^+}(\cos\theta)$ is given
by Eq.~(\ref{Eq:A_LR-cos-mf=0}),  
and is displayed in Figure~\ref{Figure:ALR-cos-ef-theta=010}.
In most of center-of-mass energy region of interest,  relations
$|Q_{e_L\mu_L}|\gg |Q_{e_L\mu_R}|$ and $|Q_{e_R\mu_R}|\gg
|Q_{e_R\mu_L}|$ are 
satisfied so that in the forward region $\cos \theta>0$, 
the LR asymmetry is approximately 
\begin{align}
A_{LR}^{f\bar{f}}(\cos\theta)
 &\simeq \frac{|Q_{e_{L} \mu_{L}}|^{2}-|Q_{e_{R} \mu_{R}}|^{2}}
{|Q_{e_{L} \mu_{L}}|^{2}+|Q_{e_{R} \mu_{R}}|^{2}}.
\end{align}

\subsection{Left-right forward-backward asymmetry}

The LR FB asymmetry
$A_{LR,FB}^{\mu^-\mu^+}(\cos\theta)$ is given by 
in Eq.~(\ref{Eq:A_LRFB-mf=0}).  It is shown in
Figure~\ref{Figure:ALRFB-ef-theta=010}. 
For $|Q_{e_L\mu_L}|\gg |Q_{e_L\mu_R}|$,
$|Q_{e_R\mu_R}|\gg|Q_{e_R\mu_L}|$,
the LR FB asymmetry can be written in terms of
the integrated LR asymmetry $A_{LR}^{\mu^-\mu^+}$ by
\begin{align}
A_{LR,FB}^{\mu^-\mu^+}(\cos\theta)&\simeq 
 \frac{2 \cos\theta}{1+\cos^2\theta}
\frac{|Q_{e_{L}\mu_{L}}|^{2}-|Q_{e_{R}\mu_{R}}|^{2}}
 {|Q_{e_{L}\mu_{L}}|^{2}+|Q_{e_{R}\mu_{R}}|^{2}}
\simeq \frac{2\cos\theta}{1+\cos^2\theta}
A_{LR}^{\mu^-\mu^+} ~.
\end{align}

\section{Summary}
\label{Sec:Summary}

We showed some results for cross sections, forward-backward (FB),
left-right (LR), and  left-right forward-backward (LR FB)
asymmetries in the process  $e^-e^+\to \mu^-\mu^+$ 
in the GUT inspired GHU model (B-model). 
Clear deviations from the SM should be observed in the early stage of
ILC 250$\,$GeV experiments. In particular, GHU predicts strong
dependence on the polarization of $e^-$ and $e^+$ beams, with which one
can explore physics at the KK mass scale of 15$\,$TeV. 

In the talk, we focused on the analysis of the $s$-channel scattering
processes $e^-e^+\to \mu^-\mu^+$ mediated by
neutral vector bosons $Z'$ in the GHU B-model.
For $e^-e^+\to e^-e^+$, there is a contribution  not only from 
the $s$-channel scattering process but also from the $t$-channel
scattering process
\cite{Abe:1994sh,MoortgatPick:2005cw,Schael:2013ita,Bardin:2017mdd,Borodulin:2017pwh,Richard:2018zhl}. 
It has been pointed out in Ref.~\cite{Richard:2018zhl}
that for the scattering process  $e^-e^+\to e^-e^+$, 
deviations from the SM in the GHU A-model can be detected even
in the early stage of the ILC experiment at 250 GeV. Therefore
we also expect similar deviations from the SM in the GHU B-model.
We are now preparing for a paper about the $e^-e^+\to e^-e^+$ scattering
process in GHU, which will appear soon.

\section*{Acknowledgments}

This work was supported in part 
by European Regional Development Fund-Project Engineering Applications of 
Microworld Physics (No.\ CZ.02.1.01/0.0/0.0/16\_019/0000766) (Y.O.), 
by the National Natural Science Foundation of China (Grant Nos.~11775092, 
11675061, 11521064, 11435003 and 11947213) (S.F.), 
by the International Postdoctoral Exchange Fellowship Program (IPEFP)
(S.F.),  
and by Japan Society for the Promotion of Science, 
Grants-in-Aid  for Scientific Research, No.\ JP19K03873 (Y.H.)
and Nos.\  JP18H05543 and JP19K23440 (N.Y.).

\appendix

\section{Observables}
\label{Sec:Observables-app}

Here we summarize formulas of several observables in the $s$-channel
scattering processes of $e^-e^+\to f\bar{f}$ mediated by only neutral
vector bosons $V_i$. In the proceeding, we consider
$f\bar{f}=\mu^-\mu^+$, so we neglect the final state fermion mass.
The formulas with nonzero fermion mass are give in e.g.,
Ref.~\cite{Funatsu:2020haj}. For more information, see
Ref.~\cite{Funatsu:2020haj} and the references.

\subsection{Cross section}

The differential cross section for the $e^-e^+\to f\bar{f}$ process
is given by 
\begin{align}
&\frac{d\sigma^{f\bar{f}}}{d\cos\theta}(P_{e^-},P_{e^+},\cos\theta) \cr
\noalign{\kern 5pt}
&\hskip 1.cm
=\left(1 - P_{e^-} P_{e^+}\right)
 \frac{1}{4}\biggl\{
 (1 - P_{\rm eff})\frac{d\sigma_{LR}^{f\bar{f}}}{d\cos\theta}(\cos\theta)
+(1 + P_{\rm eff})\frac{d\sigma_{RL}^{f\bar{f}}}{d\cos\theta}(\cos\theta)
\biggr\}
\label{Eq:dsigma_P}
\end{align}
where $P_{e^\pm}$ denotes longitudinal polarization of $e^\pm$.
$P_{e^\pm} = +1$ corresponds to purely right-handed $e^\pm$.
$P_{\rm eff}$ is defined as
\begin{align}
 P_{\rm eff}\equiv \frac{P_{e^-}-P_{e^+}}{1-P_{e^-}P_{e^+}} ~.
\label{Eq:P_eff}
\end{align}
${d\sigma_{LR}}/{d\cos\theta}$ and
${d\sigma_{RL}}/{d\cos\theta}$ are
differential cross sections for
$e_L^-e_R^+\to f\bar{f}$ and $e_R^-e_L^+\to f\bar{f}$:
\begin{align}
\frac{d\sigma_{LR}^{f\bar{f}}}{d\cos\theta}(\cos\theta)
&\simeq \frac{s}{32\pi}\biggl\{
 \left(1+\cos\theta\right)^2|Q_{e_{L} f_{L}}|^{2}  
+\left(1-\cos\theta\right)^2|Q_{e_{L} f_{R}}|^{2}  
\biggr\},
\nonumber\\
\frac{d\sigma_{RL}^{f\bar{f}}}{d\cos\theta}(\cos\theta)
&\simeq \frac{s}{32\pi}\biggl\{
 \left(1+\cos\theta\right)^2|Q_{e_{R} f_{R}}|^{2}  
+\left(1-\cos\theta\right)^2|Q_{e_{R} f_{L}}|^{2}  
\biggr\}.
\label{Eq:dsigma_LR-RL-mf=0}
\end{align}
$Q_{e_{X} f_{Y}}$ $(X,Y=L,R)$ are given by
\begin{align}
Q_{e_{X} f_{Y}}
&\equiv 
\sum_{i} \frac{g_{V_{i}e}^{X} g_{V_{i}f}^{Y} }{(s-m_{V_{i}}^{2}) + i m_{V_{i}}\Gamma_{V_{i}}},
\label{Eq:Qs}
\end{align}
where $g_{V_{i}f}^{L/R}$ are  couplings of the left- and right-handed 
fermion  $f$ to the vector boson $V_{i}$, and  $m_{V_{i}}$ and $\Gamma_{V_{i}}$ are
the mass and total decay width of $V_{i}$.

The differential cross section integrated over the angle
$\theta=[\theta_1,\theta_2]$ is given by
\begin{align}
 \sigma^{f\bar{f}}(P_{e^-},P_{e^+},[\cos\theta_1,\cos\theta_2])\equiv 
 \int_{\cos\theta_1}^{\cos\theta_2}
 \frac{d\sigma^{f\bar{f}}}{d\cos\theta}(P_{e^-},P_{e^+},\cos\theta)
 d\cos\theta,
\label{Eq:sigma-integral}
\end{align}
where
$\frac{d\sigma^{f\bar{f}}}{d\cos\theta}(P_{e^-},P_{e^+},\cos\theta)$ is 
given in Eq.~(\ref{Eq:dsigma_P}).
The observed total cross section $\sigma^{f\bar{f}}_{\rm tot}(P_{e^-},P_{e^+})$
is given by
\begin{align}
 \sigma^{f\bar{f}}_{\rm tot}(P_{e^-},P_{e^+})=
 \sigma^{f\bar{f}}
 (P_{e^-},P_{e^+},[-\cos\theta_{\rm max},+\cos\theta_{\rm max}]),
\end{align}
where the available value of $\theta_{\rm max}$ depends on each
experiment. 
The cross section $\sigma^{f\bar{f}}_{\rm tot}(P_{e^-},P_{e^+})$
is given  by
\begin{align}
\sigma_{\rm tot}^{f\bar{f}}(P_{e^-}, P_{e^+})
 &=(1 - P_{e^-} P_{e^+}) \cdot
 \frac{1}{4}   
\biggl\{
  (1 - P_{\rm eff}) \sigma_{LR}^{f\bar{f}}
 +(1 + P_{\rm eff}) \sigma_{RL}^{f\bar{f}} \biggr\},
\label{Eq:sigma_P}
\end{align}
where for the available value of $\theta_{\rm max}=1$,  
\begin{align}
\sigma_{LR}^{f\bar{f}}
 \simeq \frac{s}{12\pi}
 \left(|Q_{e_{L} f_{L}}|^{2} + |Q_{e_{L} f_{R}}|^{2}\right),\ \ \
\sigma_{RL}^{f\bar{f}}
\simeq \frac{s}{12\pi}
\left(|Q_{e_{R} f_{R}}|^{2} + |Q_{e_{R} f_{L}}|^{2}\right).
\label{Eq:sigma_LR-RL_mf=0}
\end{align}

The amount of the deviation from the SM in  the differential cross section for $e^-e^+\to f\bar{f}$
is characterized by
\begin{align}
 \Delta_{d\sigma}^{f\bar{f}}(P_{e^-},P_{e^+},\cos\theta)\equiv 
 \frac{\myfrac{d\sigma_{\rm GHU}^{f\bar{f}}}
 {d\cos\theta}(P_{e^-},P_{e^+},\cos\theta)}
 {\myfrac{d\sigma_{\rm SM}^{f\bar{f}}}{d\cos\theta}(P_{e^-},P_{e^+},\cos\theta)}-1 ~. 
\label{Eq:Delta_dsigma}
\end{align}

\subsection{Forward-backward asymmetry}

The forward-backward asymmetry $A_{FB}^{{f\bar{f}}}(P_{e^-},P_{e^+})$
\cite{Schrempp:1987zy,Kennedy:1988rt} is given by
\begin{align}
A_{FB}^{{f\bar{f}}}(P_{e^-},P_{e^+})
&=\frac{\sigma_F^{{f\bar{f}}}(P_{e^-},P_{e^+})
  -\sigma_B^{{f\bar{f}}}(P_{e^-},P_{e^+})}
 {\sigma_F^{{f\bar{f}}}(P_{e^-},P_{e^+})
 +\sigma_B^{{f\bar{f}}}(P_{e^-},P_{e^+})} ~, \cr
\noalign{\kern 5pt}
\sigma_F^{f\bar{f}}(P_{e^-},P_{e^+})&=
 \sigma^{f\bar{f}}(P_{e^-},P_{e^+},[0,+\cos\theta_{\rm max}]) ~, \cr
\noalign{\kern 5pt}
\sigma_B^{{f\bar{f}}}(P_{e^-},P_{e^+})&=
 \sigma^{f\bar{f}}(P_{e^-},P_{e^+},[-\cos\theta_{\rm max},0]) ~,
\label{Eq:A_FB}
\end{align}
By using $Q_{e_Xf_Y}  (X,Y=L,R)$ in Eq.~(\ref{Eq:Qs}),
for $\cos \theta_{\rm max} =1$,
\begin{align}
&A_{FB}^{f\bar{f}}(P_{e^-},P_{e^+}) \simeq \frac{3}{4} \frac{B_1 - B_2}{B_1 + B_2} ~, \cr
\noalign{\kern 5pt}
&\quad 
B_1 = (1+P_{\rm eff}) |Q_{e_{R} f_{R}}|^{2} + (1-P_{\rm eff})|Q_{e_{L} f_{L}}|^{2} ~, \cr
\noalign{\kern 5pt}
&\quad 
B_2 = (1+P_{\rm eff}) |Q_{e_{R} f_{L}}|^{2} + (1-P_{\rm eff})|Q_{e_{L} f_{R}}|^{2} ~, 
\label{Eq:A_FB-mf=0}
\end{align}
where $P_{\rm eff}$ is given in Eq.~(\ref{Eq:P_eff}).

The amount of the deviation from the SM is characterized by
\begin{align}
 \Delta_{A_{FB}}^{f\bar{f}}\equiv 
 \frac{A_{FB,{\rm GHU}}^{f\bar{f}}}{A_{FB,{\rm SM}}^{f\bar{f}}}-1 ~.
\label{Eq:Delta_A_FB}
\end{align}

\subsection{Left-right asymmetry}

The left-right asymmetry
\cite{Schrempp:1987zy,Kennedy:1988rt,MoortgatPick:2005cw}
is given by
\begin{align}
A_{LR}^{{f\bar{f}}}(\cos\theta)
 =  \frac{
 \sigma_{LR}^{f\bar{f}}(\cos\theta)
 -\sigma_{RL}^{f\bar{f}}(\cos\theta)
 }
 {
 \sigma_{LR}^{f\bar{f}}(\cos\theta)
 +\sigma_{RL}^{f\bar{f}}(\cos\theta)
 },
\label{Eq:A_LR-cos}
\end{align}
where
$\sigma_{LR}^{f\bar{f}}(\cos\theta)$ and
$\sigma_{RL}^{f\bar{f}}(\cos\theta)$ stand for
$\frac{d\sigma_{LR}^{f\bar{f}}}{d\cos\theta}(\cos\theta)$ and 
$\frac{d\sigma_{RL}^{f\bar{f}}}{d\cos\theta}(\cos\theta)$
in Eq.~(\ref{Eq:dsigma_LR-RL-mf=0}), respectively.
\begin{align}
A_{LR}^{f\bar{f}}(\cos\theta)
 &\simeq \frac{(1+\cos\theta)^2
 \left(|Q_{e_{L} f_{L}}|^{2}-|Q_{e_{R} f_{R}}|^{2}\right)
 +(1-\cos\theta)^2
 \left(|Q_{e_{L} f_{R}}|^{2}-|Q_{e_{R} f_{L}}|^{2}\right)
}
{(1+\cos\theta)^2
 \left(|Q_{e_{L} f_{L}}|^{2}+|Q_{e_{R} f_{R}}|^{2}\right)
 +(1-\cos\theta)^2
 \left(|Q_{e_{L} f_{R}}|^{2}+|Q_{e_{R} f_{L}}|^{2}\right)
 }.
\label{Eq:A_LR-cos-mf=0} 
\end{align}
The observable left-right asymmetry is given by
\begin{align}
A_{LR}^{{f\bar{f}}}(P_{e^-},P_{e^+},\cos\theta)
& =  \frac{
 \sigma^{f\bar{f}}(P_{e^-},P_{e^+},\cos\theta)
 -\sigma^{f\bar{f}}(-P_{e^-},-P_{e^+},\cos\theta)
 }
 {
 \sigma^{f\bar{f}}(P_{e^-},P_{e^+},\cos\theta)
 +\sigma^{f\bar{f}}(-P_{e^-},-P_{e^+},\cos\theta)
 }
\label{Eq:A_LR-cos-obs}
\end{align}
for $P_{e^-}<0$ and $|P_{e^-}|>|P_{e^+}|$,
where
$\sigma^{f\bar{f}}(P_{e^-},P_{e^+},\cos\theta)$ and 
$\sigma^{f\bar{f}}(-P_{e^-},-P_{e^+},\cos\theta)$ stand for
$\frac{d\sigma^{f\bar{f}}}{d\cos\theta}(P_{e^-},P_{e^+},\cos\theta)$ and 
$\frac{d\sigma^{f\bar{f}}}{d\cos\theta}(-P_{e^-},-P_{e^+},\cos\theta)$
in Eq.~(\ref{Eq:dsigma_P}), respectively.
(\ref{Eq:A_LR-cos-obs}) is related to (\ref{Eq:A_LR-cos}) by
\begin{align}
A_{LR}^{{f\bar{f}}}(\cos\theta)=\frac{1}{P_{\rm eff}}
A_{LR}^{{f\bar{f}}}(P_{e^-},P_{e^+},\cos\theta) ~.
\label{A_LR_cos}
\end{align}

The integrated left-right asymmetry
$A_{LR}^{f\bar{f}}$
\cite{Schrempp:1987zy,Kennedy:1988rt}
is given by
\begin{align}
 A_{LR}^{{f\bar{f}}}
 &= \frac{\sigma_{LR}^{{f\bar{f}}}-\sigma_{RL}^{{f\bar{f}}}}
 {\sigma_{LR}^{{f\bar{f}}}+ \sigma_{RL}^{{f\bar{f}}}}.
\label{Eq:A_LR}
\end{align}
By using $Q_{e_Xf_Y} (X,Y=L,R)$ in Eq.~(\ref{Eq:Qs}),
$A_{LR}^{{f\bar{f}}}$ is expressed as
\begin{align}
 A_{LR}^{{f\bar{f}}}
 &\simeq 
\frac{
[|Q_{e_{L} f_{L}}|^{2} + |Q_{e_{L} f_{R}}|^{2}]
 - [|Q_{e_{R} f_{R}}|^{2} + | Q_{e_{R} f_{L}}|^{2}] 
}{
[|Q_{e_{L} f_{L}}|^{2} + |Q_{e_{L} f_{R}}|^{2}]
 + [|Q_{e_{R} f_{R}}|^{2} + | Q_{e_{R} f_{L}}|^{2}] 
}.
\label{Eq:A_LR-mf=0}
\end{align}
The observable left-right asymmetry is given by 
\begin{align}
A_{LR}^{{f\bar{f}}}(P_{e^-},P_{e^+})
 =  \frac{
 \sigma^{f\bar{f}}(P_{e^-},P_{e^+})
 -\sigma^{f\bar{f}}(-P_{e^-},-P_{e^+})
 }
 {
 \sigma^{f\bar{f}}(P_{e^-},P_{e^+})
 +\sigma^{f\bar{f}}(-P_{e^-},-P_{e^+})
 }
\label{Eq:A_LR-obs}
\end{align}
for $P_{e^-}<0$ and $|P_{e^-}|>|P_{e^+}|$.
It is related to (\ref{Eq:A_LR}) by
\begin{align}
A_{LR}^{{f\bar{f}}}=\frac{1}{P_{\rm eff}}
A_{LR}^{{f\bar{f}}}(P_{e^-},P_{e^+}) ~.
\end{align}

The amount of the deviation from the SM in (\ref{A_LR_cos}) and (\ref{Eq:A_LR}) is 
characterized by
\begin{align}
\Delta_{A_{LR}}^{f\bar{f}}(\cos\theta) &\equiv 
\frac{A_{LR,{\rm GHU}}^{{f\bar{f}}}(\cos\theta)}{A_{LR,{\rm
 SM}}^{{f\bar{f}}}(\cos\theta)}  -1 ~,\nonumber\\
\noalign{\kern 5pt}
\Delta_{A_{LR}}^{f\bar{f}} &\equiv 
 \frac{A_{LR,{\rm GHU}}^{f\bar{f}}}{A_{LR,{\rm SM}}^{f\bar{f}}}-1 ~.
\label{Eq:Delta_A_LR}
\end{align}

\subsection{Left-right forward-backward asymmetry}

The left-right forward-backward asymmetry
\cite{Blondel:1987gp,Kennedy:1988rt,Abe:1994bj,Abe:1994bm,Abe:1995yh}
is given by
\begin{align}
A_{LR,FB}^{f\bar{f}}(\cos\theta)&=
\frac{\left[\sigma_{LR}^{f\bar{f}}(\cos\theta)
- \sigma_{RL}^{f\bar{f}}(\cos\theta)\right]
- \left[\sigma_{LR}^{f\bar{f}}(-\cos\theta)
- \sigma_{RL}^{f\bar{f}}(-\cos\theta)\right]
}{
\left[\sigma_{LR}^{f\bar{f}}(\cos\theta)
+ \sigma_{RL}^{f\bar{f}}(\cos\theta)\right]
+ \left[\sigma_{LR}^{f\bar{f}}(-\cos\theta)
+ \sigma_{RL}^{f\bar{f}}(-\cos\theta)\right]} ~.
\label{Eq:A_LRFB}
\end{align}
In terms of  $Q_{e_Xf_Y} (X,Y=L,R)$ in Eq.~(\ref{Eq:Qs}),
$A_{LR,FB}^{{f\bar{f}}}$ is expressed as
\begin{align}
A_{LR,FB}^{f\bar{f}}(\cos\theta)&\simeq
 \frac{2 \cos\theta}{1+\cos^2\theta} \frac{D_-}{D_+},
 \nonumber\\
\noalign{\kern 5pt}
D_\pm &=  \big( |Q_{e_{L} f_{L}}|^{2} + |Q_{e_{R} f_{L}}|^{2} \big) 
          \pm \big(  |Q_{e_{L} f_{R}}|^{2} + |Q_{e_{R} f_{R}}|^{2} \big)  ~. 
\label{Eq:A_LRFB-mf=0}
\end{align}
The observable left-right forward-backward asymmetry is given by
\begin{align}
 &A_{LR,FB}^{{f\bar{f}}}(P_{e^-},P_{e^+},\cos\theta) = \frac{E_-}{E_+}~,
 \nonumber\\
&E_\pm =  \big[  \sigma^{f\bar{f}}(P_{e^-},P_{e^+},\cos\theta) + \sigma^{f\bar{f}}(-P_{e^-},-P_{e^+},-\cos\theta) \big] \cr
\noalign{\kern 5pt}
&\hskip 1.2cm
\pm  \big[ \sigma^{f\bar{f}}(-P_{e^-},-P_{e^+},\cos\theta) +\sigma^{f\bar{f}}(P_{e^-},P_{e^+},-\cos\theta) \big] 
\label{Eq:A_LRFB-cos-obs}
\end{align}
for $P_{e^-}<0$ and $|P_{e^-}|>|P_{e^+}|$.
The relation between $A_{LR,FB}^{{f\bar{f}}}(\cos\theta)$ in
Eq.~(\ref{Eq:A_LRFB}) and
$A_{LR,FB}^{{f\bar{f}}}(P_{e^-},P_{e^+},\cos\theta)$ in 
Eq.~(\ref{Eq:A_LRFB-cos-obs}) is given by
\begin{align}
A_{LR,FB}^{{f\bar{f}}}(\cos\theta)=\frac{1}{P_{\rm eff}}
A_{LR,FB}^{{f\bar{f}}}(P_{e^-},P_{e^+},\cos\theta) ~. 
\end{align}

The amount of the deviation in $A_{LR,FB}$ from the SM is characterized by
\begin{align}
 \Delta_{A_{LR,FB}}^{f\bar{f}}(\cos\theta)\equiv 
 \frac{A_{LR,FB,{\rm GHU}}^{f\bar{f}}(\cos\theta)}
{A_{LR,FB,{\rm SM}}^{f\bar{f}}(\cos\theta)} -1 ~.
\label{Eq:Delta-A_LRFB}
\end{align}

{\small
\bibliographystyle{utphys} 
\bibliography{../../arxiv/reference}
}

\end{document}